\documentclass[12pt,letterpaper]{article}

\usepackage{amsmath}
\usepackage{amsfonts}
\usepackage{bm}        
\usepackage{dsfont}    
\usepackage{mathrsfs}  
\usepackage{fullpage}
\usepackage[colorlinks,linkcolor=blue,citecolor=blue]{hyperref}
\usepackage{graphicx}
\usepackage{titlesec}

    \newcommand{\e}{\mathrm{e}}                
    \newcommand{\eqn}[1]{Eq. (\ref{#1})}       
    \renewcommand{\vec}[1]{\bm{#1}}            
    \newcommand{\nvec}[1]{\vec{\hat{#1}}}      
    \newcommand{\group}[1]{\mathsf{#1}}        
    \newcommand{\algb}[1]{\mathfrak{#1}}       
    \newcommand{\tr}{\mathrm{tr}}              
    \renewcommand{\span}{\mathrm{span}}        
    \renewcommand{\O}{\mathcal{O}}             
    \newcommand{\SU}[1]{\group{SU}(#1)}        
    \newcommand{\su}[1]{\algb{su}(#1)}         
    \newcommand{\tH}{\tilde{H}}                

    \def\>{\rangle}                            
    \def\<{\langle}                            %
    \renewcommand{\i}{\mathrm{i}}              
    \newcommand{\Id}{\mathds{1}}               
    \renewcommand{\H}[1]{H_{\mathrm{#1}}}
    \newcommand{\Hx}{\H{x}}                    
    \newcommand{\Hy}{\H{y}}                    
    \newcommand{\Hz}{\H{z}}                    
    \renewcommand{\u}{\vec{u}}
    \newcommand{\uu}[1]{u_{\mathrm{#1}}}
    \newcommand{\du}{\vec{\delta u}}
    \newcommand{\ux}{\uu{x}}
    \newcommand{\uy}{\uu{y}}
    \newcommand{\uz}{\uu{z}}
    \newcommand{\iprod}[2]{\<#1,#2\>}          
    \newcommand{\hsnorm}[1]{||#1||_{\mathrm{HS}}} 
    \newcommand{\F}{\mathcal{F}}               
    \newcommand{\ep}[1]{\epsilon_{#1}}
    \newcommand{\p}[1]{\phi_{\mathrm{#1}}}

    \newcommand{\U}{U}               
    \newcommand{\R}{R}

\begin{document}




\titleformat{\section}{\large \bfseries}{\thesection}{1em}{}
\titleformat{\subsection}{\normalsize \bfseries}{\thesubsection}{1em}{}
\titleformat{\subsubsection}{\normalsize \bfseries}{\thesubsubsection}{1em}{}
\renewcommand{\contentsname}{}

\begin{center}
\thispagestyle{empty}
\uppercase{ {\large \bf Progress in compensating pulse sequences for quantum computation }}\\
\vspace{1em}
\uppercase{  J. True Merrill}\footnote{ Email: true.merrill@gatech.edu} \uppercase{and Kenneth R. Brown} \\
{\small \today}\\

\vspace{1em}
\emph{Schools of Chemistry and Biochemistry; Computational Science and
    Engineering; and Physics, Georgia Institute of Technology, Atlanta,
    GA 30332, USA}
\end{center}
\vspace{1em} 

The control of qubit states is often impeded by systematic control
errors.  Compensating pulse sequences have emerged as a resource
efficient method for quantum error reduction.  In this review, we
discuss compensating composite pulse methods, and introduce a unifying
control-theoretic framework using a dynamic interaction picture.  This
admits a novel geometric picture where sequences are interpreted as
vector paths on the dynamical Lie algebra.  Sequences for single-qubit
and multi-qubit operations are described with this method.

\vspace{-3em}
\tableofcontents


\section{Introduction}

In any experiment, external noise sources and control errors limit the
accuracy of the preparation and manipulation of quantum states.  In
quantum computing, these effects place an important fundamental limit
on the size and accuracy of quantum processors.  These restrictions
may be reduced by quantum error correction.  Although very
sophisticated quantum error-correcting codes exist which are robust
against any general error, these codes require large-scale
multipartite entanglement and are very challenging to implement in
practice \cite{Gottesman2010}.  Therefore, it is of great interest to
investigate schemes which reduce errors with a smaller resource
overhead.  One alternative strategy involves replacing an error-prone
operation by a pulse sequence which is robust against the error.

The basis of our strategy is that all noises and errors can be treated
as an unwanted dynamic generated by an error Hamiltonian. This error
Hamiltonian can arise either through interactions with the environment
or by the misapplication of control fields. This view unifies the
pulse sequences developed for combating unwanted interactions, e.g.,
dynamic decoupling and dynamically corrected gates \cite{Viola1999,
    Khodjasteh2005, Khodjasteh2009}, with those for overcoming
systematic control errors, compensating composite pulse sequences
\cite{Levitt1986}.  In each case, the methods are limited by the rate
at which control occurs relative to the time scale over which the
error Hamiltonian fluctuates. However, many experiments are limited by
control errors and external fields that vary slowly relative to the
time scale of a single experimental run but vary substantially over
the number of experiments required to obtain precise results.

In this review paper, we examine a number of techniques for handling
unwanted control errors including amplitude errors, timing errors, and
frequency errors in the control field.  We emphasize the common
principles used to develop compensating pulse sequences and provide a
framework in which to develop new sequences, which will be of use to
both quantum computation and coherent atomic and molecular
spectroscopy.


\section{Coherent control over spin systems}

The accurate control of quantum systems is an important prerequisite
for many applications in precision spectroscopy and in quantum
computation.  In complex experiments, the task reduces to applying a
desired unitary evolution using a finite set of controls, which may be
constrained by the physical limitations of the experimental apparatus.
Stimulated by practical utility, quantum control theory has become an
active and diverse area of research \cite{Huang1983, Schirmer2002,
    Albertini2003, Mabuchi2005, D'Alessandro}.  Although originally
developed using nuclear-magnetic resonance (NMR) formalism,
compensating pulse sequences can be approached from the perspective of
quantum control theory with unknown systematic errors in the controls
\cite{Li2006}. Here we review several fundamental concepts in quantum
control, and apply these ideas using NMR as an instructive example.
Although we restrict the discussion to control in NMR spectroscopy,
the following analysis is quite general as several other coherent
systems (e.g. semiconductor quantum dots \cite{Taylor2007},
superconducting qubits \cite{Chiorescu2003, Martinis2003, Clarke2008},
and trapped ions \cite{Leibfried2003, Haffner2008}) may be considered
by minor modifications to the Hamiltonian.

In practice, a desired evolution is prepared by carefully manipulating
the coupling of the system to a control apparatus, such as a
spectrometer.  In the absence of relaxation, the coherent dynamics are
governed by the quantum propagator $U(t)$, which in non-relativistic
quantum mechanics must satisfy an operational Schr\"odinger equation,
\begin{eqnarray}
\dot{U}(t) = -\i \left ( \sum_\mu u_\mu(t) H_\mu \right ) U(t), \qquad U(0) = \Id.
\label{eq:control}
\end{eqnarray}
In this model, the unitless Hamiltonians $H_\mu \in \{H_1, H_2, \dots
, H_n \}$ are modulated by real-valued control functions $u_\mu(t) \in
\{ u_1(t), u_2(t), \dots, u_n(t) \}$ and represent the $n$ available
degrees of control for a particular experimental apparatus.  In
analogy with linear vector spaces, we interpret the vector $\u(t) =
(u_1(t), u_2(t), \dots , u_n(t))$ as a vector function over the
manifold of control parameters, with components $u_\mu(t)$
representing the magnitudes of the control Hamiltonians with units of
angular frequency.  It is convenient to introduce a second vector of
control Hamiltonians $\vec{H} = (H_1, H_1, \dots , H_n)$ and the short-hand
notation $H(t) = \u(t) \cdot \vec{H}$.  We omit a term which represents the
portion of the total Hamiltonian which is outside of direct control
(i.e. a drift Hamiltonian); in principle it is always possible to work
in an interaction picture where this term is removed.  Alternatively,
one may assign a Hamiltonian $H_0$ to represent this interaction, with
the understanding that $u_0(t) = 1$ for all $t$.

For a given control system, a natural question concerns the optimal
approximation of a desired unitary propagation using a set of
constrained control functions.  Constraints may include limitations on
the total operation length, control amplitudes or derivatives.  The
study of this question requires the solution of \eqn{eq:control} for a
particular set of controls; such solutions may be obtained using
several methods, including the Dyson series \cite{Dyson1949}, and the
Magnus \cite{Magnus1954, Blanes2009}, Fer \cite{Madhu2006}, and Wilcox
\cite{Wilcox1967} expansions.  We label particular solutions to the
control equation over the interval $t_i \leq t \leq t_f$ as $U(\u(t);
t_f, t_i)$.  If the set of all possible solutions to \eqn{eq:control}
is the set of all unitary gates on the Hilbert space (i.e. the
solutions form a representation of $\group{U}(n)$ or $\SU{n}$, to be
discussed in section \ref{lie}) then the system is \emph{operator
    controllable} \cite{Albertini2003a, D'Alessandro}.

\begin{figure}
\begin{center}
\includegraphics{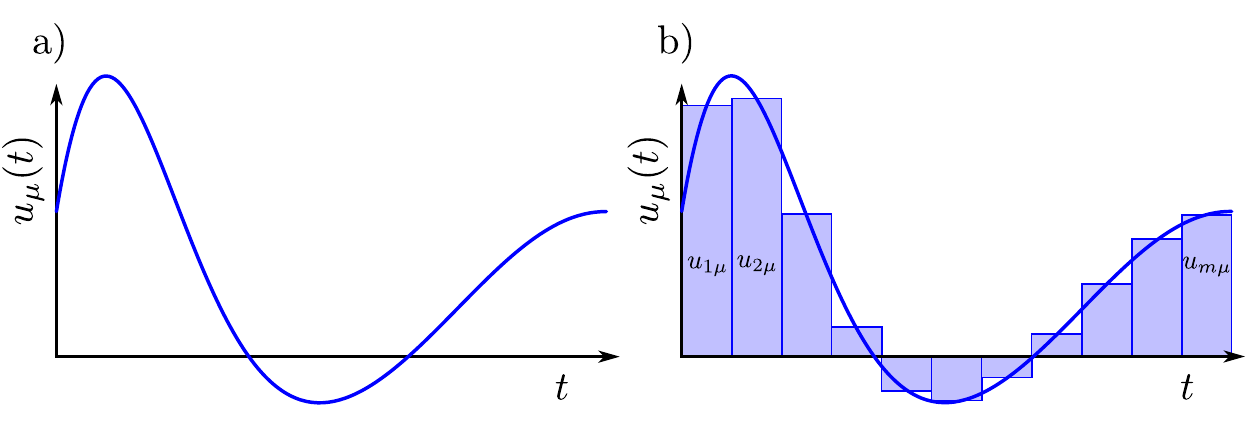}
\end{center}
\caption{The experimental controls available to manipulate a quantum
    system are modeled using a set of real-valued control functions
    $\{u_\mu(t)\}$ that modify a set of dimensionless Hamiltonians
    $\{H_\mu\}$.  a) An example control function $u_\mu(t)$. b) A
    discrete approximation for $u_\mu(t)$ composed of square
    pulses. \label{fig:control}}
\end{figure}

For some applications, it is convenient to assume that the operation
time $\tau$ is discretized into $m$-many time intervals, over which
the control functions $u_\mu$ are constant, i.e., during the $k$th
time interval $\Delta t_k$ the Hamiltonian is time-independent and the
resulting evolution operator is $U_k = U(\u_{k}; t_k + \Delta t_k ,
t_k ) = \exp( -\i \Delta t_k \u_k \cdot \vec{H} )$.  Figure
\ref{fig:control} illustrates an example control function and a
possible discretization scheme.  Over each time interval, the applied
unitary operation is a square pulse.  In many experiments, the
application of gates using sequences of square pulses is preferred for
simplicity.  The discretization of the control functions into square
pulses allows one to solve \eqn{eq:control} in a piecewise fashion.
The total propagator for a sequence of $m$-many time steps is given by
the time-ordered product
\begin{eqnarray}
U(\u(t),\tau,0) = U_m U_{m-1} \dots U_{k} \dots U_{2}U_{1} = \prod_{k=1}^{m} U_{k},
\end{eqnarray}
where the multiplication of each successive operator is understood to
be taken on the left; this is in agreement with standard quantum
mechanics conventions, where operations are ordered right-to-left, but
at odds with some NMR literature where successive operations are
ordered left-to-right.  Frequently we will consider propagators over
the entire duration of a sequence $0 \leq t \leq \tau$; as a matter of
notational convenience we drop the time interval labels whenever there
is no risk of confusion.  If a pulse sequence $U(\u(t))$ is equivalent
to a target operation $U_T$ in the sense that during an experiment,
$U(\u(t))$ may be substituted for $U_T$, then $U(\u(t))$ is a
\emph{composite pulse sequence} \cite{Levitt1981, Freeman1998}.

The usefulness of composite pulse sequences lies in that in many cases,
one may simulate a target unitary transformation $U_T$, which may be
difficult to directly implement, by instead implementing a sequence of
simpler pulses which under some set of conditions is equivalent to
$U_T$.  Composite pulse sequences may be designed to have several
important advantages over a directly applied unitary, such as improved
resilience to errors. The properties of pulses or pulse sequences may
be considered either by the transformations produced on a particular
initial state $\rho$, or by comparing the sequence to an ideal
operation, which contains information on how the sequence transforms
all initial states. Following Levitt \cite{Levitt1986}, we assign
composite pulses into two classes: the fully-compensating class A, and
the partially-compensating class B. The properties of these classes
are briefly reviewed.

\emph{Class A:} All composite pulse sequences in class A may be
written in the form
\begin{eqnarray}
\e^{\i \phi} U(\u(t)) = U_T,
\label{eqn:fully-compensating}
\end{eqnarray}
where it is assumed that the individual pulses in the sequence are
error-free.  The global phase $\phi$ is irrelevant to the dynamics
since for any initial state $\rho$, both the pulse sequence and the
target operation apply the same transformation ($U(\u(t)) \rho
U^{\dagger}(\u(t)) = U_T \rho U_T^\dagger$).  Sequences in this class
are suited for use in quantum computation, since the transformation is
independent of the initial quantum state.  The study of these
sequences will be the primary topic of this article.

\emph{Class B:} Composite pulse sequences in class B transform one
particular initial condition to a set of final conditions, which for
the purposes of the experiment, are equivalent.  For example, consider
an NMR experiment on a spin $I = 1/2$ nucleus where the nuclear
magnetization, initially oriented ``spin-up'' (i.e. $\rho = \Id/2 +
\Hz$, where $\Hx$, $\Hy$, $\Hz$ are the nuclear angular momentum
operators) is transferred to the $\Hx$-$\Hy$ plane by a pulse
sequence.  In this sense, all sequences that apply the transformations
are equivalent up to a similarity transform $\exp( -\i \beta \Hz)$,
which applies a $\Hz$ phase to the spin.  Other sequences in class B
may satisfy $U(\u(t)) \rho U^{\dagger}(\u(t)) = U_T \rho U_T^\dagger$
for a particular initial state $\rho$, but fail to satisfy
\eqn{eqn:fully-compensating}.  All sequences in class B are generally
not well suited for use in quantum computation, since implementation
requires specific knowledge of the initial and final states of the
qubit register.  Class B sequences are however very useful in other
applications, including NMR \cite{Levitt1979, Levitt1981,
    Freeman1998}, MRI \cite{DeGraaf2005}, control over nitrogen
vacancy centers \cite{Said2009} and in ion trapping experiments
\cite{Schmidt-Kaler2003a}.  We close our discussion by noting that
class B sequences may be converted into a fully-compensating class A
sequence by a certain symmetrical construction \cite{Luy2005}.


\subsection{Errors in quantum control}
\label{errors}
We now consider the effects of unknown errors in the control
functions.  Recall that a pulse sequence may be specified by a set of
control functions $\{u_\mu(t)\}$, which we group into a control vector
$\u(t)$. Suppose however, during an experiment an unknown systematic
error deforms each of the applied controls from $u_\mu(t)$ to
$v_\mu(t)$.  In the presence of unknown errors, the perfect
propagators $U(\u(t);t_f,t_i)$ are replaced with their imperfect
counterparts $V(\u(t);t_f,t_i) = U(\vec{v}(t);t_f,t_i)$, which may be
regarded as an image of the perfect propagator under the deformation
of the controls.

\begin{figure}
\begin{center}
\includegraphics{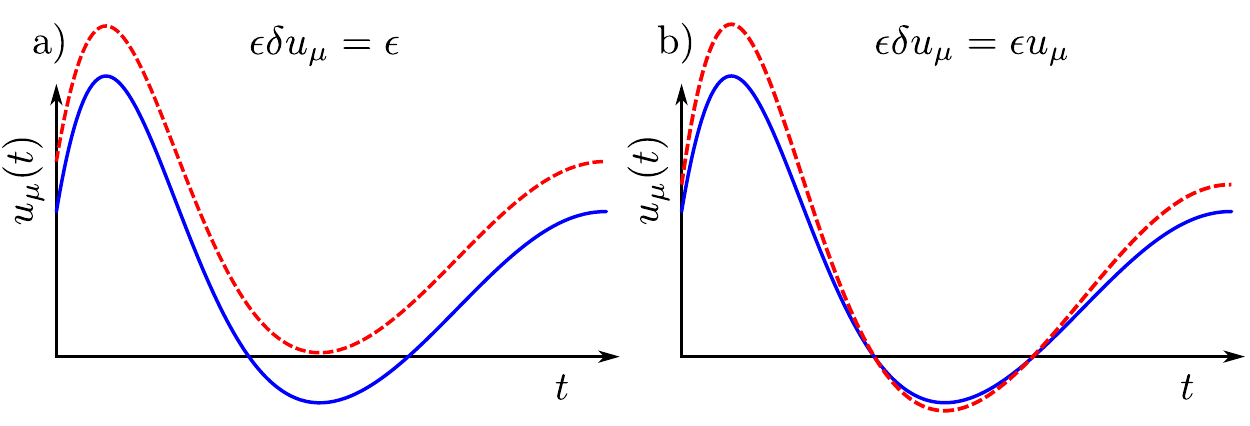}
\end{center}
\caption{Systematic errors induce deformations of the ideal control
    functions $u_\mu(t)$ (solid curves) to the imperfect controls
    $v_\mu(t) = u_\mu(t) + \epsilon \delta u_\mu(t)$ (dashed curves).
    Common error models include a) a constant unknown offset in the
    control, and b) an error in the amplitude of the control
    function. \label{fig:error-control}}
\end{figure}

In practice, systematic errors typically arise from a miscalibration
of the experimental system, for example an imprecise measurement of the
intensity or frequency of a controlling field.  In these cases, it is
appropriate to introduce a deterministic model for the control
deformation.  Specifically, $\vec{v}(t)$ must be a function of the
controls $\u(t)$, and for each component of the imperfect control
vector we may write
\begin{eqnarray}
  v_\mu(t) = f_\mu[ \vec{u}(t); \epsilon ],
\end{eqnarray}
where the functional $f_\mu$ is the \emph{error model} for the
control and the variable $\epsilon$ is an unknown real parameter that
parametrizes the magnitude of the error.  This construction may be
generalized to the case of multiple systematic errors by considering
error models of the form $f_\mu[\u(t);\epsilon_i,\epsilon_j, \dots,
\epsilon_k]$.  The model $f_\mu$ is set by the physics of the problem,
and is chosen to produce the correct evolution under the imperfect
controls.  Formally, we may perform the expansion
\begin{eqnarray}
 f_\mu[ \u(t); \epsilon ] = f_\mu[ \u(t); 0 ] + \epsilon \frac{d}{d \epsilon} f_\mu[ \u(t); 0] + \frac{\epsilon^2}{2!} \frac{d^2}{d \epsilon^2} f_\mu[ \u(t); 0 ] + \O(\epsilon^3),
\end{eqnarray}
then by the condition that when $\epsilon = 0$ the control must be
error free, it is trivial to identify $u_\mu(t) = f_\mu[ \u(t) ; 0 ]$.
Frequently, it is sufficient to consider models which are linear in
the parameter $\epsilon$.  In this case we introduce the shorthand
notation $\delta u_\mu(t) = \frac{d}{d\epsilon} f_\mu[ \u(t) ; 0 ]$
and the corresponding vector $\du(t)$ to represent the first-order
deformation of the controls, so that the imperfect controls take the
form $v_\mu(t) = u_\mu(t) + \epsilon \delta u_\mu(t)$.  Figure
\ref{fig:error-control} illustrates two common error models: constant
offsets from the ideal control value ($f_\mu[\u(t);\epsilon] =
u_\mu(t) + \epsilon$), and errors in the control amplitude
($f_\mu[\u(t);\epsilon] = (1+\epsilon) u_\mu(t)$).

A natural question to ask is what effect unknown errors have on the
evolution of the system.  It is obvious that imperfect pulses make
accurate manipulation of a quantum state difficult.  One may be
surprised to find that for some cases, the effects of errors on the
controls may be systematically removed, without knowledge of the
amplitude $\epsilon$.  The method we describe involves implementing a
compensating composite pulse sequence which is robust against
distortion of the controls by a particular error model.  As an
example, consider a case where an experimentalist would like to
approximate a the target unitary $U_T = U(\vec{u}(t);\tau,0)$, where
at least one of the controls is influenced by a systematic error.  The
target operation may be simulated up to $\O(\epsilon^n)$ if there
exists a set of control functions $\u(t)$ such that,
\begin{eqnarray}
V(\u(t);\tau,0) = U(\u(t) + \epsilon \du(t);\tau,0) = U(\u(t);\tau,0) +  \O(\epsilon^{n+1}).
\label{eq:compensation-condition}
\end{eqnarray}
There are many (infinite in most cases) sets of control functions
$\u(t)$ which implement a target unitary transformation, however only
a small subset of possible control functions are robust to distortion
by a particular systematic error.  If robust controls can be found,
then pulse sequence $V(\u(t);\tau,0)$ can be applied in the place of
$U_T$, and the leading order terms of the offset $\epsilon \du(t)$ are
suppressed.  A sequence with these properties is called a
\emph{compensating pulse sequence}, and may be thought of as a set of
control functions which are optimized to remove the effect of
leading-order terms of unknown systematic errors in the controls.  By
construction, sequences of this form are fully compensating (Class A).

When an experimentalist implements a compensating pulse sequence they
attempt to apply the ideal operations $U(\u(t);\tau,0)$ ignorant of
the amplitude $\epsilon$ of a systematic error.  However, the
operations are not ideal and to emphasize this we introduce
$V(\u(t);\tau,0)$ to represent the \emph{imperfect propagators} that
is actually implemented when $U(\u(t);\tau,0)$ is attempted.  The
functional dependence of the error in $V(\u(t);\tau,0)$ is not
explicitly written allowing us to study different error models with
the same pulse sequence.

Let us consider the dynamics of the system under the interaction frame
Hamiltonian $H^I(t) = \sum_\mu \epsilon \delta u_\mu(t) H^{I}_\mu
(t)$, where $H^{I}_\mu (t) = U^\dagger(\u(t');t,0) H_\mu
U(\u(t');t,0)$ are the control Hamiltonians in the interaction frame.
In this picture, $H^I(t)$ is regarded as a perturbation, and we
associate the propagator $U^I(\du (t); \tau, 0)$ as the particular
solution to the interaction picture Schr\"odinger equation over the
interval $0 \leq t \leq \tau$.  Hence
\begin{eqnarray}
V( \u(t); \tau , 0) = U(\u(t); \tau, 0) U^I(\epsilon \du(t); \tau, 0),
\label{eq:interaction-definition}
\end{eqnarray}
and from \eqn{eq:compensation-condition}
\begin{eqnarray}
 U^I(\epsilon \du (t);\tau,0) = \Id + \O(\epsilon^{n+1}).
\label{eq:interaction-compensation}
\end{eqnarray}
Quite generally, when a fully compensating pulse sequence is
transformed into the interaction frame the resulting propagator must
approximate the identity operation \cite{Ichikawa2011b}.  The
techniques for constructing compensating pulse sequences discussed in
the present article rely on performing a series expansion by powers of
$\epsilon$ for the interaction frame propagator $U^{I}(\epsilon
\du(t);\tau,0)$, then choosing a set of controls which remove the
leading terms of the distortion, and finally transforming back out of
the interaction frame.


\subsection{NMR spectroscopy as a model control system}
\label{nmr}
In this section, we will apply the ideas developed thus far to a model
one-qubit NMR quantum computer \cite{Vandersypen2005, Jones2011},
which serves as a relevant example of a system where coherent control
is possible.  In section \ref{two-qubit} multi-qubit operations are
considered.  There are many possible physical implementations for a
qubit, oftentimes based on a two-level subsystem of a larger Hilbert
space.  In this case, the qubit is defined on the angular momentum
states of a spin-$1/2$ nucleus.

\begin{figure}
\begin{center}
\includegraphics[width = 4 in]{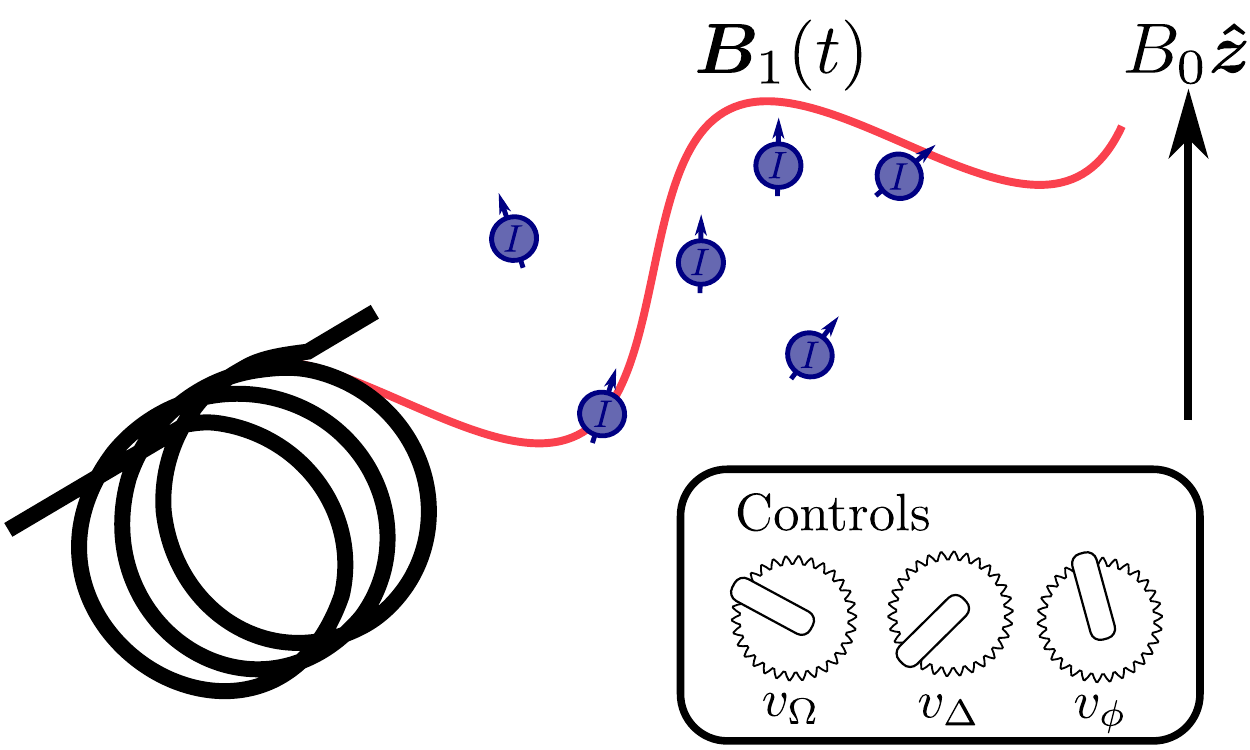}
\end{center}
\caption{Simplified diagram of an NMR spectrometer.  Control over an
    ensemble of nuclear spins is applied by a radiofrequency (rf)
    field $\vec{B}_1(t)$, where the Rabi frequency $\Omega(t)$, field
    detuning $\Delta(t)$, and phase $\phi(t)$ are control
    parameters. Unknown systematic control errors (e.g. poor intensity
    control, detuning errors) may be present. \label{fig:NMR}}
\end{figure}

Consider an ensemble of spin $I = 1/2$ nuclei undergoing Larmor
precession under a static magnetic field $\vec{B}_0$ oriented along
the $\nvec{\mathrm{z}}$ axis.  The static field $\vec{B}_0$ induces a
net magnetization among the nuclei.  A transverse radiofrequency (rf)
field near nuclear resonance $\vec{B}_1(t) = \vec{B}_1 \cos(\omega_1 t
- \phi)$ is applied in the $\nvec{\mathrm{x}}$-$\nvec{\mathrm{y}}$
plane.  A simplified diagram of an NMR spectrometer is provided in
figure \ref{fig:NMR}.  The analysis is simplified by assuming that
individual nuclei in the ensemble are decoupled by the rapid tumbling
of spins in the sample \cite{Freeman1998}.  After transforming into the
rotating frame, the Hamiltonian for a single spin may be written as
\begin{eqnarray}
\label{eq:nmr}
H(t) =  \Delta(t) \Hz + \Omega(t) \big ( \cos (\phi(t)) \Hx + \sin (\phi(t)) \Hy  \big ), 
\end{eqnarray}
where
\begin{eqnarray*}
\Hx = \frac{1}{2} \begin{pmatrix} 0&1 \\ 1&0 \end{pmatrix}, \qquad \Hy = \frac{1}{2} \begin{pmatrix} 0&-\i 
\\ \i&0 \end{pmatrix}, \qquad \Hz = \frac{1}{2} \begin{pmatrix} 1&0 \\ 0&-1 \end{pmatrix},
\end{eqnarray*}
and the counter-rotating terms have been neglected under the
rotating-wave approximation.  Manipulation of the rf field provides a
convenient set of controls to guide the evolution of the spins; it is
assumed that the rf field detuning $\Delta(t)$, phase $\phi(t)$, and
the Rabi frequency $\Omega(t)$ are each independently controllable and
may suffer from independent systematic errors.  Let the controls
$\vec{u}(t)$ denote the vector $(\Omega(t) \cos(\phi(t)), \Omega(t)
\sin(\phi(t)), \Delta(t))$.  Then the Hamiltonian may be written as
$H(t) = \u(t) \cdot \vec{H}$, and the resulting Schr\"{o}dinger
equation for the propagator is of the form of \eqn{eq:control}.  Thus,
the coherent spin dynamics in magnetic resonance spectroscopy can be
reformulated in terms of a problem in quantum control.


\subsubsection{Error models}
\label{nmr-error-models}
In practice, systematic errors in the controls caused by instrumental
limitations prohibit the application of perfect pulse propagators.  We
consider several models for errors in the controls of a one-qubit NMR
quantum computer.  Often compensating pulse sequences are
well optimized for one type of error, but provide no advantage against
a different error model.

\emph{Amplitude errors:} An amplitude error arises from slow
systematic variation in the amplitude of the rf field, resulting in a
small offset in the applied Rabi frequency.  Let $\Omega$ represent
the ideal Rabi frequency and $\Omega'$ represent the offset.  From the
form of the control vector $\u(t)$ it follows that in the presence of
the error, the $\ux(t)$ and $\uy(t)$ controls are distorted, i.e. they
are replaced by their imperfect counterparts $v_\mathrm{x/y}(t) =
\uu{x/y}(t) + \ep{A} \delta \uu{x/y}(t)$, where $\delta \uu{x/y}(t) =
\uu{x/y}(t)$ is proportional to the ideal control value and the error
parameter $\ep{A} = \Omega' / \Omega < 1$ is the relative amplitude of
the field offset.  The imperfect pulses take the form $V(\u(t)) =
U(\u(t)(1+\ep{A}))$.

\emph{Pulse length errors:} A pulse length error is a systematic error
in the duration of individual pulses, perhaps due to an offset in the
reference oscillator frequency.  The imperfect propagator takes the
form $V(\vec{u}(t); t_0 + \Delta t, t_0) = U(\vec{u}(t); t_0 + \Delta
t + \delta t, t_0)$, where $\Delta t$ is the ideal pulse length and
$\delta t$ is the unknown timing error.  In some cases, errors on the
clock may be rewritten in terms of equivalent errors on the control
functions.  For simplicity, we restrict ourselves to square pulses,
where the controls $\vec{u}(t)$ are constant and the imperfect
propagator may be rewritten as $V(\vec{u}; t_f, t_i) = U(\vec{u} +
\ep{T} \du; t_0 + \Delta t, t_0)$, where again $\du = \u$ is
proportional to the ideal control and now $\ep{T} = \delta t / \Delta
t$, similar to the result for an amplitude error.  Amplitude and pulse
length errors in square pulses are similar in some senses, though they
arise from distinct physical processes, since they both act as errors
in the angle of the applied rotation.

\emph{Addressing errors:} Individual spins among the ensemble may
experience slightly different Rabi couplings owing to the spatial
variation in the strength of the control field.  In some applications
this variation is exploited to yield spatially localized coherent
operations, such as in magnetic resonance imaging, and in addressing
single atoms in optical lattices or single ions in ion trap
experiments \cite{Ivanov2011, Haffner2008}.  In fact, many proposed
scalable architectures for quantum processors rely on this effect to
discriminate between qubits.  In these cases, it is important to
distinguish between the evolution applied to the addressed spins and
the evolution of spins outside of the addressed region, where ideally
no operation is applied.

Consider an experiment where the ensemble of spins is divided into a
high field, $\Omega$, and low field, $\Omega' < \Omega$, region.  We
assume that the control of the addressed (high-field) spins is perfect
(i.e., for addressed spins $V(\u(t)) = U(\u(t))$); however, the spins
in the low-field region experience an undesired correlated rotation
$V(\u(t)) = U(\ep{N} \u(t))$, where $\ep{N} = \Omega' / \Omega$ (the
subscript $N$ denotes neighboring spins).  In many cases, the
imperfect pulses on the unaddressed qubits may be regarded as very
small rotations.  From a mathematical point of view, sequences
composed of these rotations are more easily attacked using the
expansion techniques that will be developed in section \ref{BCH}.

\emph{Detuning errors:} Systematic errors may also arise in the
control of the frequency of the rf field.  Consider the case where
an experimentalist attempts to perform an operation at a particular
field tuning $\Delta(t)$, however a slow unknown frequency drift
$\delta(t)$ is present.  From the form of the NMR controls, it follows
that the error distorts the $\Hz$ component to $v_\mathrm{z}(t) =
\uz(t) + \ep{D} \delta \uz$, where $\uz = \Delta(t)$, and $\ep{D}
\delta \uz(t) = \delta(t)$.  Unlike the other models discussed thus
far, the detuning error applies a shift in the effective rotation
axis, rotating it in the direction of the $\Hz$ axis.  Therefore, the
ideal propagator and the imperfect counterpart do not commute in
general.


\subsection{Binary operations on unitary operators}
In our discussion of the properties of compensating pulse sequences,
it will be necessary to study the structure of the control
Hamiltonians $\{H_\mu\}$, and to calculate the accuracy of gates.
Here we discuss the Hilbert-Schmidt product and the fidelity measure.

\emph{Hilbert-Schmidt inner product:} The Hilbert-Schmidt inner
product (also known as the Frobenius product) is a natural extension
of the vector inner product over the field of complex numbers to
matrices with complex coefficients.  Let $U$ and $V$ be $n \times n$
matrices with entries $U_{ij}$ and $V_{ij}$ respectively.  The
Hilbert-Schmidt inner product $\iprod{U}{V}$ is defined as,
\begin{eqnarray}
\iprod{U}{V} = \sum_{i=1}^n \sum_{j=1}^n U^{*}_{ji} V_{ij} = \tr(U^\dagger V).
\end{eqnarray}
This inner product has several properties that closely mirror the
inner product for complex vectors, namely $\iprod{U}{V} =
\iprod{V}{U}^*$, and $|\iprod{U}{V}| \leq \hsnorm{U} \: \hsnorm{V}$,
where the Hilbert-Schmidt norm $\hsnorm{U} = \sqrt{ \iprod{U}{U} }$
also satisfies a corresponding triangle inequality $\hsnorm{U+V} \leq
\hsnorm{U} + \hsnorm{V}$.  
This strong correspondence to Euclidean vector spaces will be
exploited in section \ref{lie}, where Lie-algebraic techniques are
employed to construct compensating pulse
sequences. 

\emph{Fidelity:} A useful measure for evaluating the effects of a
systematic control error is the operational fidelity, defined as
\begin{eqnarray}
\F(U,V) = \min_{\psi} \sqrt{ \<\psi|U^\dagger V|\psi\>\<\psi|V^\dagger U|\psi\>}.
\end{eqnarray}
For any two unitary matrices in the group $\SU{2}$, the fidelity may
be written as $\F(U,V) = |\iprod{U}{V}|/2$. In this paper, we show
error improvement graphically by plotting the infidelity, $1-\F(U,V)$.


\section{Group theoretic techniques for sequence design} In the
present work, we briefly discuss several aspects of the theory of Lie
groups which are useful for constructing compensating pulse sequences,
with emphasis on conceptual clarity over mathematical rigor.  The
interested reader is referred to Refs. \cite{D'Alessandro} and
\cite{Gilmore} for additional details and a rigorous treatment of the
subject.  Quantum control theory and the related study of continuous
transformation groups is a rich and active subject, with applications
in chemistry \cite{Brumer1986, Demiralp1993, Ramakrishna1995}, and
other applications requiring the precise manipulation of quantum
states \cite{Khaneja2001.2, Khaneja2002}.  We now turn our attention
to continuous groups of quantum transformations, specifically
transformations which induce qubit rotations in quantum computing.


\subsection{Lie groups and algebras}
\label{lie}
Consider the family of unitary operators that are solutions to the
control equation
\begin{eqnarray}
\label{eq:control2}
\dot{U}(t) =  \sum_\mu u_\mu(t) \tH_\mu U(t), \qquad U(0) = \Id,
\end{eqnarray}
where $\tH_\mu = -\i H_\mu$ are skew-symmetrized Hamiltonians with
 the added condition that the (possibly infinite) set of control
Hamiltonians $\{\tH\}$ is closed under the commutation operation (in
the sense that for all $\tH_\mu, \: \tH_\nu \in \{\tH\}$, $\alpha
[\tH_\mu,\tH_\nu] \in \{\tH\}$, where $\alpha \in \mathds{R}$ ).  
This extra condition does not exclude any of the ``physical'' operators 
which are
generated by a subset of $\{\tH\}$, since these operators correspond
to solutions where the remaining control functions $u_\mu(t)$ have
been set to zero.  Observe that these solutions form a representation
of a group, here denoted as $\group{G}$, since the
following properties are satisfied: for all solutions $U_1$ and $U_2$,
the product $U_1U_2$ is also a solution (see section \ref{BCH}); the
associative property is preserved (i.e. $U_1(U_2U_3) = (U_1U_2)U_3$);
the identity $\Id$ is a valid solution; and for all
solutions $U_1$, the inverse $U_1^\dagger$ is also a valid solution.
Moreover, the group forms a continuous differentiable manifold,
parametrized by the control functions.  A continuous group which is
also a differentiable manifold with analytic group multiplication and
group inverse operations is called a \emph{Lie group}.  We identify
the group $\group{G}$ of solutions as a Lie group, and note that
differentiation of the elements $U \in \group{G}$ is well defined by
nature of \eqn{eq:control2}.

We now turn our attention to elements of $\group{G}$ in the
neighborhood of the identity element, that is, infinitesimal unitary
operations.  In analogy with differentiation on Euclidean spaces, note
that for any $U(t) \in \group{G}$ one may find a family of tangent
curves at $U(0) = \Id$,
\begin{eqnarray}
 \left. \frac{dU(t)}{dt} \right |_{t=0} = \sum_\mu u_\mu(0) \tH_\mu.
\end{eqnarray}
The set of skew-symmetrized Hamiltonians $\{ \tH \}$ and the field of
real numbers $\mathds{R}$ (corresponding to the allowed values for the
components $u_\mu(0)$) form a linear vector space under matrix
addition and the Hilbert-Schmidt inner product \cite{Gilmore}.  Here,
the Hamiltonians $\tH$ take the place of Euclidean vectors and span
the tangent space of $\group{G}$ at the identity, denoted by
$T_\Id\group{G}$.  A homomorphism exists between the control functions
$\u(0)$ and vectors in $T_\Id\group{G}$.  On this space is defined the
binary Lie bracket operation between two vectors $[\tH_\mu ,
\tH_\nu]$, which for our purposes is synonymous with the operator
commutator.  A vector space which is is closed under the Lie bracket
is an example of a \emph{Lie algebra}.  By construction, the set $\{
\tH \}$ is closed under commutation and therefore forms a Lie algebra,
here denoted as $\algb{g}$, corresponding to the Lie group
$\group{G}$.  We note that in general the Hamiltonians may be linearly
dependent; however, an orthogonal basis under the Hilbert-Schmidt
product may be generated using an orthogonalization algorithm.

The power of most Lie algebraic techniques relies on the mapping
between group elements which act on a manifold (such as the manifold
of rotations on a Bloch sphere) to elements in a Lie algebra which are
members of a vector space.  In the groups we study here, the mapping
is provided by the exponential function $\group{G} = \e^{\algb{g}}$
(i.e., every element $U \in \group{G}$ may be written as $U = \e^{g}$
where $g \in \algb{g}$).  The object of this method is to study the
properties of composite pulse sequences, which are products of members
of a Lie group, in terms of vector operations on the associated Lie
algebra.


\subsubsection{The spinor rotation group  $\SU{2}$}
\label{su2-example}
As a relevant example, consider the group of single-qubit operations
generated by the NMR Hamiltonian \eqn{eq:nmr}.  This is a
representation of the special unitary group $\SU{2}$.  The
skew-symmetrized control Hamiltonians are closed under the Lie bracket
and thus form a representation of the Lie algebra $\su{2} = \span
\{-\i \Hx, -\i \Hy, -\i \Hz\}$ \footnote{The operation $\span$ denotes
    all linear combinations with real coefficients}.  Therefore, any
element in $U \in \SU{2}$ may be written as $U = \e^{ -\i t \u \cdot
    \vec{H} }$, where $-\i t \vec{u} \cdot \vec{H} \in \su{2}$ may now
be interpreted as a vector on the the Lie algebra.  Furthermore, since
$\iprod{-\i H_\mu}{-\i H_\nu} = \delta_{\mu,\nu} / 2$, the spin
operators form an orthogonal basis for the algebra.  Topologically
$\SU{2}$ is compact and is homomorphic to rotations of the 2-sphere,
the group $\group{SO}(3)$ \cite{Gilmore}; there are exactly two group
elements $U$ and $-U \in \SU{2}$ which map to the same rotation.


\subsection{Baker-Campbell-Hausdorff and Magnus formulas}
\label{BCH}
A composite pulse sequence may be studied from the perspective of
successive products between elements of a Lie group.  In the following
analysis, it will be useful to relate the product of two members of a
Lie group to vector operations on the Lie algebra.  The relationship
allows us to map pulse sequences to effective Hamiltonians.  This
correspondence is provided by the Baker-Campbell-Hausdorff (BCH)
formula \cite{Gilmore, Grensing1986}, which relates group products to
a series expansion in the Lie algebra.  For rapid convergence, it is
most convenient to consider products of infinitesimal unitary
operations, that is, operations of the form $\e^{\epsilon g}$, where
$g \in \algb{g}$ and $\epsilon < 1$ is a real expansion parameter.  We
assume $\epsilon$ is sufficiently small to guarantee that the group
product of propagators always lies within the radius of convergence
for the expansion \cite{Blanes2009}.  Let $U_1 = \e^{\epsilon \tH_1}$
and $U_2 = \e^{\epsilon \tH_2}$ be members of a Lie group $\group{G}$,
where the Hamiltonians $\tH_1$, $\tH_2 \in \algb{g}$ are members of
the associated algebra. The BCH representation for the product $U_1U_2
= U_3$ involves the calculation of an effective Hamiltonian $\tH_3 \in
\algb{g}$ by the expansion
\begin{eqnarray}
U_3 = \exp( \tH_3 ) =  \exp \left( \sum_n^{\infty} \epsilon^n F_n  \right),
\end{eqnarray}
where the terms
\begin{eqnarray*}
\epsilon F_1 &=& \epsilon (\tH_1 + \tH_2) \\
\epsilon^2 F_2 &=& \frac{\epsilon^2}{2}[\tH_1,\tH_2] \\
\epsilon^3 F_3 &=& \frac{\epsilon^3}{12} \Big([\tH_1,[\tH_1,\tH_2]] + [\tH_2,[\tH_2,\tH_1]] \Big),
\end{eqnarray*}
are calculated from $\tH_1$, $\tH_2$, and nested commutators of
elements of the Lie algebra.  A combinatoric formula found by Dykin
\cite{Dykin1947} exists to calculate $F_n$ for arbitrary $n$.  The
expansion may be truncated once a desired level of accuracy is
reached.  In principle, group products of arbitrarily length may be
approximated to arbitrary accuracy using BCH formulas; however, these
formulas rapidly become unwieldy and difficult to use without the aid
of a computer. The BCH expansion is most useful for sequences of
square pulses, where for each pulse the Hamiltonian is time
independent.

\emph{The Magnus expansion:} 
\label{magnus}
A related expansion developed by Magnus \cite{Magnus1954} may be used
to compute the propagator generated by a general time-dependent
Hamiltonian.  The solution to a control equation (e.g. $\dot{U}(t) =
\epsilon \tH(t) U(t)$, where $\epsilon \tH(t) = \sum_\mu \epsilon
\delta u_\mu(t) \tH_\mu$) over the interval $t_i \leq t \leq t_f$ may
be written as the power series
\begin{eqnarray}
U(\epsilon \du;t_f,t_i) = \exp \left( \sum_n^\infty \epsilon^n \Omega_n(t_f,t_i) \right),
\label{eq:magnus}
\end{eqnarray}
where the first few expansion terms are,
\begin{eqnarray*}
\epsilon \Omega_1(t_f,t_i) &=& \epsilon \int_{t_i}^{t_f} dt \tH(t) \\
\epsilon^2 \Omega_2(t_f,t_i) &=& \frac{\epsilon^2}{2} \int_{t_i}^{t_f} dt \int_{t_i}^{t} dt' [\tH(t),\tH(t')] \\
\epsilon^3 \Omega_3(t_f,t_i) &=& \frac{\epsilon^3}{6} \int_{t_i}^{t_f} dt \int_{t_i}^{t} dt' \int_{t_i}^{t'} dt'' ([\tH(t),
[\tH(t'), \tH(t'')]] + [\tH(t''),[\tH(t'), \tH(t)]]).
\end{eqnarray*}
Again as a matter of notational convenience, we drop the time interval
labels $(t_f,t_i)$ on the expansion terms when there is no risk of
confusion.  Formulas for higher order terms may be found in
\cite{Burum1981}.  The BCH and Magnus expansions are in fact
intimately related; when considering piecewise-constant controls the
techniques are equivalent. We refer the interested reader to
\cite{Blanes2009} for further details regarding both the Magnus
expansion and BCH formulas.

The BCH and Magnus expansions are very well known in composite pulse
literature, and techniques that utilize these expansions are
collectively referred to as average Hamiltonian theory. A variant of
this technique, pioneered by Waugh \cite{Haeberlen1968, Waugh2007},
has been a mainstay of composite pulse design in the NMR community for
decades.  In some formalisms, the BCH expansion of the product of two
propagators is interpreted as a power series in the rotation axis and
angle.  For the study of composite single-qubit rotations, this
picture is extremely useful as it allows rotations on the sphere to
guide the mathematics.  However, this picture of composite rotations
can not be generalized to more complex groups, such as the group of
$n$-qubit operations $\SU{2^n}$.

In this work, we emphasize a Lie algebraic interpretation of these
methods, which also leads to a second geometric picture for the terms
of the expansions.  Observe that the first-order terms $F_1$
and $\Omega_1$ may be regarded as simple vector sums on the
Lie algebra, i.e., the sum of $\tH_1$ and $\tH_2$ in the BCH expansion,
and the sum of each of the infinitesimal vectors $\tH(t)dt$ in the
Magnus expansion.  In an analogous way, one may interpret the higher
order terms as the addition of successively smaller vectors on
$\algb{g}$.  From this insight, one may construct composite sequences
which simulate a target unitary from geometric considerations on the
Lie algebra.

\begin{figure}
\begin{center}
\includegraphics{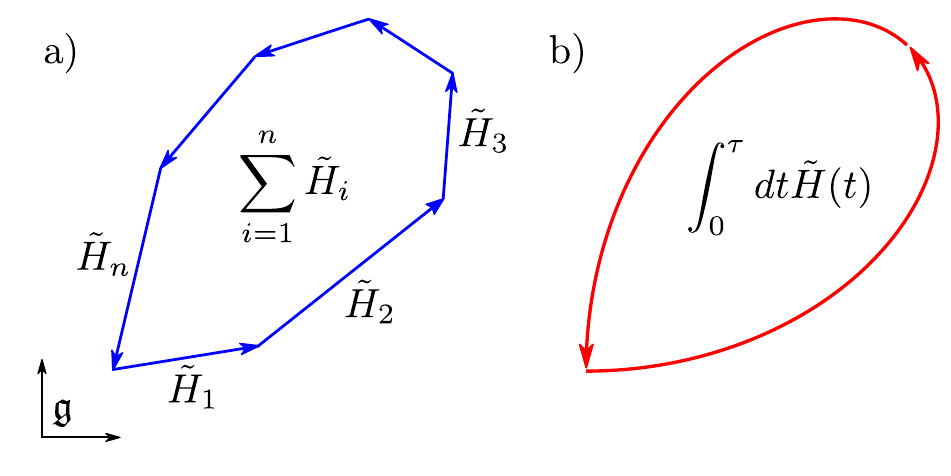}
\end{center}
\caption{Vector paths on a Lie algebra $\algb{g}$ may be used to
    represent a pulse sequence. a) The BCH expansion relates group
    multiplication of several square pulses to vector addition on
    $\algb{g}$.  b) A shaped pulse is represented by a vector curve on
    $\algb{g}$, parametrically defined by the control functions.  Both
    pulse sequences form a closed loop on $\algb{g}$ and therefore the
    first-order terms equal zero.}
\label{cycle}
\end{figure}


\subsubsection{A method for studying compensation sequences} 
\label{method}
At this point, it is useful to introduce the general method that will
be used to study compensating pulse sequences.  Recall from section
\ref{errors} that in the presence of an unknown systematic control
error, the ideal control functions are deformed into imperfect
analogues.  The imperfect propagator may be decomposed as
$V(\u(t);\tau,0) = U(\u(t);\tau,0) U^I(\epsilon \du(t);\tau,0)$, where
$U^I(\epsilon \du(t);\tau,0)$ represents the portion of the evolution
produced by the systematic distortion of the ideal controls.  Provided
that the displacements $\epsilon \du(t)$ are sufficiently small
relative to the ideal controls, we may perform a Magnus expansion for
the interaction frame propagator
\begin{eqnarray}
U^I(\epsilon \du(t);\tau,0) = \exp \left( \sum_{n=1}^{\infty} \epsilon^n \Omega_n(\tau,0) \right ),
\label{eq:magnus-cancellation}
\end{eqnarray}
where the integrations in the expansion terms are performed in the
appropriate frame.  For example, the first-order term is
\begin{eqnarray}
\epsilon \Omega_1(\tau,0) = -\i \epsilon \int_0^\tau dt \:  \du (t) \cdot  \vec{H}^I(t),
\label{eq:first-order}
\end{eqnarray}
where the components $H^I_\mu(t) = U^\dagger(\u(t');t,0) H_\mu
U(\u(t');t,0)$ are the interaction frame control Hamiltonians.  The
reader should recall from \eqn{eq:interaction-compensation} that if
the controls $\u(t)$ form an $n$th order compensating pulse sequence,
then the interaction frame propagator $U^I(\epsilon \du(t);\tau,0) =
\Id + \O(\epsilon^{n+1})$ must approximate the identity to sufficient
accuracy.  It immediately follows that this condition is satisfied for
any $\epsilon$ when the leading $n$-many Magnus expansion terms
$\Omega_1(\tau,0), \Omega_2(\tau,0), \dots , \Omega_n(\tau,0)$ over
the pulse interval $0 \leq t \leq \tau$ simultaneously equal zero.

This condition may also be understood in terms of geometric properties
of vector paths on the Lie algebra.  For example, consider the
first-order Magnus expansion term for the interaction frame propagator
and observe that $-\i \epsilon \du(t) \cdot \vec{H}^I(t)$ may be regarded
as a vector path on the dynamical Lie algebra.  From the condition
$\Omega_1(\tau,0) = 0$ and \eqn{eq:first-order}, we see that the path
must form a closed cycle in order for the first-order term to be
eliminated (see figure \ref{cycle}).  The elimination of higher-order
expansion terms will place additional geometric constraints on the
path which will depend on the structure of the Lie algebra (i.e. the
commutators between paths on the algebra).  In the case of piecewise
constant control functions, the resulting propagator may be understood
as a sequence of square pulses.  It is clear that in this case the
Magnus series reduces to a BCH expansion for the total pulse
propagator, and on the Lie algebra the corresponding path forms a
closed polygon. 

This Lie theoretic method is a useful tool in determining whether a
control function $\u(t)$ is also a compensating sequence; we may
directly calculate the interaction frame Magnus expansion terms in a
given error model and show that they equal zero.  The inverse problem
(i.e. solving for control functions) is typically much more difficult.
In general, the interaction Hamiltonians $H_\mu^I(t)$ are highly
nonlinear functions of the ideal controls $\u(t)$, which impedes
several analytical solution methods.  However, we note that in several
special cases the problem is considerably simplified, such as when the
ideal operation is to perform the identity and the interaction and Schr\"odinger
pictures are equivalent.


\subsection{Decompositions and approximation methods}
Several useful techniques in sequence design involve decompositions
that may be understood in terms of the structure of a Lie group and
its corresponding algebra.  In this section we discuss several important
methods and identities.


\subsubsection{Basic building operations}
Given a limited set of controls $\{\tH_1,\tH_2\}$ that generate the
algebra $\algb{g}$, one may produce any unitary operation in the
corresponding Lie group $\group{G} = \e^\algb{g}$ using only two
identities.  The first identity, the Lie-Trotter formula
\cite{Trotter1958}, describes how to produce a unitary generated by
the sum of two non-commuting control operators.  Using the BCH formula
one may compute that $\e^{\tH_1 / n} \e^{\tH_2 / n} = \e^{(\tH_1 +
    \tH_2)/n} + \O([\tH_1, \tH_2] / n^2)$.  In terms of physical
pulses, this corresponds to dividing the propagators $\e^{\tH_1}$ and
$\e^{\tH_2}$ into $n$ equal intervals to produce the propagators
$\e^{\tH_1/n}$ and $\e^{\tH_2/n}$.  Suppose we perform $n$ such
successive products, so that the resulting propagator is
\begin{eqnarray}
\left ( \e^{\tH_1/n} \e^{\tH_2/n} \right )^n = \e^{\tH_1 + \tH_2} + \O([\tH_1,\tH_2]/n).
\label{eq:trotter}
\end{eqnarray}
Although the Hamiltonians $\tH_1$ and $\tH_2$ do not commute in
general, we may approximate $U = e^{\tH_1 + \tH_2}$ to arbitrary
accuracy by dividing the evolution into $n$-many time intervals and
using the construction \eqn{eq:trotter}.  By extension, it follows
that any unitary generated by a Hamiltonian in the Lie algebra
subspace $\span\{\tH_1, \tH_2\}$ may be approximated to arbitrary
accuracy using a Trotter sequence.  A number of improved sequences
were developed by Suzuki \cite{Suzuki1992} that remove errors to
higher commutators and scale more strongly with $n$. The
Trotter-Suzuki formulas may be used to eliminate successively higher
order errors at the cost of increased operation time \cite{Wiebe2010}.

The second identity, which we refer to as the balanced group
commutator, enables the synthesis of a unitary generated by the Lie
bracket $[\tH_1 , \tH_2]$.  Again the BCH formula may be used to show
$\e^{\tH_1 / n} \e^{\tH_2 / n} \e^{-\tH_1 / n} \e^{-\tH_2 / n} =
\e^{[\tH_1 , \tH_2] / n^2} + \O( [\tH_1 + \tH_2, [\tH_1,\tH_2]] /
n^3)$.  If we now consider $n^2$-many successive balanced group
commutator constructions, the resulting propagator is
\begin{eqnarray}
\left( \e^{\tH_1 / n} \e^{\tH_2 / n} \e^{-\tH_1 / n} \e^{-\tH_2 / n} \right)^{n^2} =
\e^{[\tH_1 , \tH_2]} + \O( [\tH_1 + \tH_2, [\tH_1,\tH_2]] / n).
\end{eqnarray}
Then, as in the case of the Trotter formula, we may approximate $U =
\e^{[\tH_1, \tH_2]}$ to arbitrary accuracy by increasing the number of
intervals $n$.  Since by assumption the entire Lie algebra may be
generated by nested Lie brackets between the Hamiltonians $\tH_1$ and
$\tH_2$, this implies that any $U \in \group{G}$ may be produced by a
combination of balanced group commutator and Trotter formulas.
However, we emphasize that in almost all cases much more efficient
constructions exist.  The balanced group commutator construction also
forms the basis of the Solovay-Kitaev theorem \cite{Dawson2006}, an
important result regarding the universality of a finite gate set in
quantum computation.  Later, we will study the Solovay-Kitaev method
(see \ref{SK-method}) which produces compensating sequences of
arbitrary accuracy by using a balanced group commutator.  Specifically
we will use the formula
\begin{eqnarray}
  \exp({\tH_1 \epsilon^{k}}) \exp({\tH_2 \epsilon^{l}}) \exp({-\tH_1 \epsilon^{k}}) \exp({-\tH_2 \epsilon^{l}}) = \exp 
  ( [\tH_1,\tH_2] \epsilon^{k+l} ) + \O(\epsilon^{k+l+1}),
\label{SK}
\end{eqnarray}  
where the parameter $\epsilon < 1$ will represent the strength of a
systematic error.


\subsubsection{Euler decomposition}
\label{euler}
In section \ref{su2-example} it was shown that any one-qubit operation
$U \in \SU{2}$ may be written in the form $U = \exp( -\i t \u \cdot
\vec{H} )$.  It is well known that an alternative representation
exists, namely the Euler decomposition
\begin{eqnarray}
U = \exp(-\i \alpha_3 \Hx ) \exp(-\i \alpha_2 \Hy ) \exp(-\i \alpha_1 \Hx),
\label{eq:euler}
\end{eqnarray}
which is given by sequential rotations by the angles $\{\alpha_1,
\alpha_2, \alpha_3 \}$ about the $\Hx$, $\Hy$, and $\Hx$ axes of the
Bloch sphere.  The Euler decomposition gives the form of a pulse
sequence that produces any arbitrary one-qubit gate.  
The Euler decomposition
also gives a method of producing rotations generated by a Hamiltonian
outside of direct control.  For example, at perfect resonance $U =
\exp(-\i \theta \Hz)$ cannot be directly produced; however, the Euler
decomposition $U = \exp(-\i \frac{\pi}{2} \Hx) \exp(-\i \theta \Hy)
\exp(\i \frac{\pi}{2} \Hx)$ may be implemented.


\subsubsection{Cartan decomposition}
\label{cartan}
Let $\algb{g}$ be a semi-simple Lie algebra that may be decomposed
into two subspaces $\algb{g} = \algb{k} \oplus \algb{m}$, $\algb{m} =
\algb{k}^\perp$ satisfying the commutation relations,
\begin{eqnarray}
[\algb{k},\algb{k}] \subseteq \algb{k}, \qquad [\algb{m},\algb{k}] \subseteq \algb{m}, \qquad [\algb{m},\algb{m}] \subseteq \algb{k}.
\end{eqnarray}
Such a decomposition is called a Cartan decomposition of $\algb{g}$
\cite{D'Alessandro}.  Suppose for the moment there exists a subalgebra
$\algb{a}$ of $\algb{g}$ which is in a subspace of $\algb{m}$.  Since
$\algb{a}$ is an algebra of its own right, it is closed under the Lie
bracket $[\algb{a},\algb{a}] \subseteq \algb{a}$.  However, note
$\algb{a} \subseteq \algb{m}$ implies that $[\algb{a},\algb{a}]
\subseteq [\algb{m},\algb{m}] \subseteq \algb{k}$.  Since the
subspaces $\algb{k}$ and $\algb{m}$ are mutually orthogonal, then
$[\algb{a},\algb{a}] = \{0\}$ and the subalgebra $\algb{a}$ must be
abelian.  A maximal abelian subalgebra $\algb{a} \subseteq \algb{m}$
for a Cartan decomposition pair $(\algb{k},\algb{m})$ is called a
Cartan subalgebra \cite{D'Alessandro}.

For brevity, we state without proof an important theorem regarding the
decomposition of an operator in a group $\group{G}$ with a Lie algebra
admitting a Cartan decomposition.  Consider a Lie algebra $\algb{g}$
with a Cartan subalgebra $\algb{a}$ corresponding to the decomposition
pair $(\algb{k},\algb{m})$.  Every $U$ in the group $\group{G} =
\e^{\algb{g}}$ may be written in the form,
\begin{eqnarray}
U = K_2 A K_1,
\label{eq:kak}
\end{eqnarray}
where $K_1, K_2 \in \e^{\algb{k}}$ and $A \in \e^{\algb{a}}$.  This is
called the $KAK$ Cartan decomposition for the group $\group{G}$.  The
interested reader is referred to \cite{Gilmore} for additional
details regarding the $KAK$ decomposition.

As a relevant example, here we show how the Euler decomposition for a
propagator $U \in \SU{2}$ is a special case of a $KAK$ decomposition.
The algebra is spanned by the orthogonal basis matrices $\su{2} =
\span\{-\i\Hx,-\i\Hy,-\i\Hz\}$.  Observe that $\algb{k} = \span\{-\i
\Hx\}$ and $\algb{m} = \span\{-\i\Hy,-\i\Hz\}$ form a Cartan
decomposition for $\su{2}$.  The maximal abelian subalgebra of
$\algb{m}$ is one-dimensional; we choose $\algb{a} = \span\{-\i
\Hy\}$ although the choice $\algb{a}' = \span\{-\i \Hz\}$ would serve
just as well (i.e. the different axis conventions of the Euler
decomposition differ in the choice of a maximal abelian subalgebra).
Then by \eqn{eq:kak}, every element $U \in \SU{2}$ may be expressed in
the form $U = \exp(-\i \alpha_3 \Hx) \exp(-\i \alpha_2 \Hy) \exp(-\i
\alpha_1 \Hx)$, where the parameters $\alpha_j$ are real.  This is a
restatement of \eqn{eq:euler}, thus completing the proof.  The $KAK$
decomposition is an existence theorem, and does not provide a direct
method for the calculation of the required rotation angles $\alpha_j$.

The Cartan decomposition has important implications for
universality. For instance, if one may generate any unitary operation
over the subgroups $\e^{\algb{k}}$ and $\e^{\algb{a}}$, then any gate
in the larger group $\e^\algb{g}$ may be produced.  Similarly, if
compensation sequences exist for operations in these subgroups, then
they may be combined in the $KAK$ form to yield a compensating pulse
sequence for an operation in the larger group. Another important
application is the decomposition of large Lie groups into products of
more simple ones.  Of special interest to quantum computing is the
inductive decomposition of $n$-qubit $\SU{2^n}$ gates into products of
one-qubit $\SU{2}$ and two-qubit $\SU{4}$ rotations
\cite{Khaneja2001}.


\section{Composite pulse sequences on $\SU{2}$}
\label{SU2}
In this section, we study pulse sequences which compensate
single-qubit operations, which form a representation of the group
$\group{SU}(2)$.  Our approach is to use Lie theoretic methods to study the
effects of composite rotations.  Topologically, sequences of
infinitesimal rotations may be interpreted as paths in the
neighborhood of the group identity.  It is frequently easier to
construct sequences of infinitesimal paths on the Lie algebra, and
then map these sequences to the manifold of group operations.  Several
techniques will be used to construct compensating pulse sequences of
arbitrary accuracy, including techniques that use Solovay-Kitaev
methods and Trotter formulas.


\subsection{Solovay-Kitaev sequences}
The Solovay-Kitaev sequences, so named because the construction of the
higher order sequences involves the identity \eqn{SK} used by Kitaev
in his proof of universal control from finite gate sets
\cite{Dawson2006}, are among the simplest families of
fully-compensating composite pulse sequences. The SK sequences were
first introduced by Brown, Harrow, and Chuang \cite{Brown2004},
and are designed to compensate pulse length, amplitude, and addressing
errors using resonant pulses.  Here we show that the SK family of
sequences may be derived using the method outlined in section
\ref{method}

\emph{Narrowband behavior:} Narrowband composite pulse sequences apply
a spin rotation over only a narrow range of strengths of the control
field \cite{Wimperis1994}.  Therefore, they are most suited for
correcting addressing errors \cite{Ivanov2011}, i.e., situations where
the spatial variation in the field strength is used to discriminate
between spins in an ensemble.  The operations on the addressed qubits are 
assumed to be
error-free, whereas on the unaddressed qubits the imperfect pulses
take the form $V(\u(t)) = U(\ep{N} \u(t))$, where $\ep{N} < 1$ is the
systematic addressing error amplitude (see section
\ref{nmr-error-models}).  Narrowband sequences are a means of applying
a target operation $U_T$ on the addressed spins while removing the
leading effects of the operation on the unaddressed spins.  For a
composite pulse sequence to exhibit $n$th-order narrowband behavior,
we require that two conditions must be satisfied: first that for the
unaddressed qubits the sequence $V(\u(t))$ approximates the identity
up to $\O(\ep{N}^{n+1})$, and second, that on the targeted spins the
desired operation is applied without error.

From a mathematical perspective, systematic addressing errors are
among the easiest to consider.  Note that for the unaddressed qubits
the ideal values for the controls $\u(t)$ is zero (ideally no
operation takes place).  This implies $H^I_\mu(t) = H_\mu$, and that
the Schr\"odinger and interaction frames in
\eqn{eq:interaction-compensation} are identical.  We may develop a
sequence which corrects the control distortion without the added
complication of the passage into an interaction picture.  For this
reason, we first study narrowband sequences before considering other
error models.

We now show that a sequence with first-order narrowband properties may
be constructed using the method described in section \ref{method}.  We
consider sequences composed of three square pulses
(i.e., piecewise constant control functions) which produce the ideal
propagation,
\begin{eqnarray}
  U(\u(t)) = \prod_{k=1}^3 U_k, \qquad U_k = \exp (-\i t_k\u_k \cdot \vec{H} ).
\label{eq:3-pulse-form}
\end{eqnarray}
On an addressed qubit, $U(\u(t))$ is implemented, whereas on the
unaddressed qubits $V(\u(t)) = U(\ep{N} \u(t))$ is applied.  We may
use either a BCH or Magnus expansion to compute the applied operation
on the unaddressed qubit.  From \eqn{eq:first-order} and the error model
$v_{\mathrm{x/y}}(t) = \ep{N} u_{\mathrm{x/y}}(t)$, the first-order term is
\begin{eqnarray} 
  \ep{N} \Omega_1 = - \i \ep{N} ( t_3\u_3+t_2\u_2 +t_1\u_1 )\cdot \vec{H}.
\end{eqnarray}
The spin operators $-\i H_\mu \in \{ -\i \Hx, -\i \Hy, -\i \Hz \}$
form an orthogonal basis for $\su{2}$, and the terms $-\i \ep{N} t_k
\u_{k}\cdot \vec{H}$ may be regarded as vectors on the dynamical Lie
algebra.  In order to eliminate the first-order Magnus term $\ep{N}
\Omega_1$, the sum of the components must equal zero, that is, the
vectors must form a closed triangular path.  Such paths may be found
using elementary geometric methods.

\begin{figure}
\begin{center}
\includegraphics{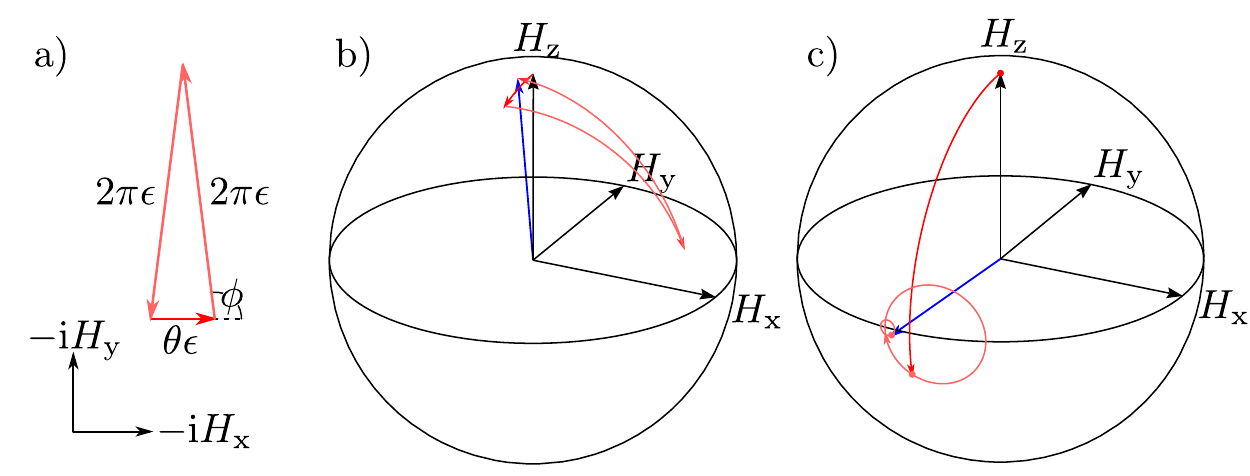}
\end{center}
\caption{a) Vector path followed by SK1 on the Lie algebra.  b)
    Trajectory of an unaddressed spin during an SK1 sequence, using
    imperfect rotations of the form $M(\theta,\phi) = R(\theta
    \ep{N},\phi)$.  c) SK1 correcting an amplitude error, using
    imperfect $M(\theta, \phi) = R(\theta(1 + \ep{A}), \phi)$. In
    these plots $\ep{N} = \ep{A} = 0.2$. \label{fig:SK1}}
\end{figure}

It is important at this point to allow experimental considerations to
place constraints on the sequences under study.  For instance, in many
cases it is desirable to perform coherent operations at resonance, a
condition which forces the vectors to lie in the $\Hx$-$\Hy$ plane.
One possible choice is
\begin{eqnarray}
  -\i \ep{N} t_1\u_{1} \cdot \vec{H} &=& -\i \ep{N} \theta \Hx \nonumber \\
  -\i \ep{N} t_2\u_{2} \cdot \vec{H} &=& -\i \ep{N} 2 \pi ( \cos \p{SK1} \Hx + \sin \p{SK1} \Hy ) \nonumber \\
  -\i \ep{N} t_3\u_{3} \cdot \vec{H} &=& -\i \ep{N} 2 \pi ( \cos \p{SK1} \Hx -\sin \p{SK1} \Hy ),
  \label{eq:SK1-generators}
\end{eqnarray}
where the phase $\p{SK1} = \arccos( - \theta / 4\pi )$ is selected so
that $\ep{N} \sum_k t_k\u_{k} = 0$, and therefore the first-order
expansion term $\ep{N} \Omega_1 = 0$ is eliminated.  Figure
\ref{fig:SK1}a is a diagram of these vectors on $\su{2}$, where the
sequence may be represented as a closed isosceles triangle with one
segment aligned on the $-\i \Hx$ axis.

For clarity, we use the notation $R(\theta, \phi) = \exp( -\i \theta (
\cos \phi \Hx + \sin \phi \Hy))$ to represent propagators which induce
rotations about an axis in the $\Hx$-$\Hy$ plane,.  Observe that for
resonant square-pulse operations in $\SU{2}$, $R(\theta,\phi) =
U(\u_k; t_k,0)$, where $\u_k = \Omega ( \cos \phi , \sin \phi , 0 )$
and $\theta = t_k |\u_k|$.  We also define the corresponding imperfect
propagator $M(\theta, \phi) = V(\vec{u};t_k,0)$, and recall that the
imperfect propagators on the unaddressed qubits $M(\theta,\phi) =
R(\theta \ep{N}, \phi)$ and addressed qubits $M(\theta,\phi) =
R(\theta,\phi)$ have different implied dependencies on the systematic
error.  Then combining \eqn{eq:3-pulse-form} and
\eqn{eq:SK1-generators}, the propagator may be written as
\begin{eqnarray}
\label{eq:SK1-unaddressed}
U(\ep{N} \u(t)) &=& R(2\pi \ep{N}, -\p{SK1}) R(2\pi \ep{N},\p{SK1}) R(\theta \ep{N}, 0) = \Id + \O(\ep{N}^2) \nonumber \\
M_\mathrm{SK1}(\theta,0) &=& M(2\pi , -\p{SK1}) M(2\pi,\p{SK1}) M(\theta,0),
\end{eqnarray}
for the unaddressed qubits, and as 
\begin{eqnarray}
\label{eq:SK1-addressed}
U(\u(t)) &=& R(2\pi,-\p{SK1}) R(2\pi, \p{SK1}) R(\theta,0) \nonumber \\
M_\mathrm{SK1}(\theta,0) &=& M(2\pi,-\p{SK1}) M(2\pi,\p{SK1}) M(\theta,0),
\end{eqnarray} 
for addressed qubits. This is the first-order Solovay-Kitaev sequence,
here denoted as SK1.  On the addressed qubits the rotations
$R(2\pi,\p{SK1}) = R(2\pi,-\p{SK1}) = -\Id$ are resolutions of the
identity and the effect of the sequence is to apply the target unitary
$U_T = R(\theta,0)$, whereas on the unaddressed spins the sequence
applies $\Id + \O(\ep{N}^2)$ and the leading first-order rotation of
the unaddressed spins is eliminated.  Therefore, SK1 satisfies the
conditions for narrowband behavior.  When the sequence SK1 is used in
the place of the simple rotation $R(\theta, 0)$, the discrimination
between addressed and unaddressed spins is enhanced.

\emph{Broadband behavior:} Broadband composite pulses apply a spin
rotation over a large range of strengths of the control field, and are
best suited for correcting systematic amplitude and pulse-length
errors that correspond to systematic
over/under-rotations during qubit manipulations.  Broadband sequences
are a means of applying a target operation in the presence of
inaccurate field strengths or pulse durations.  We require for a
sequence to exhibit $n$th-order broadband behavior, the effect of the
sequence is to approximate the target operation $U_T$ up to
$\O(\ep{A}^n)$ in the case of amplitude errors, or up to
$\O(\ep{T}^n)$ in the case of pulse-length errors.  As before, it is
convenient to consider sequences comprised of resonant square pulses.
In this case the amplitude and pulse length error models are in some
sense equivalent since they both apply a proportional distortion to
the $\ux(t)$ and $\uy(t)$ controls.  In the following discussion we
will explicitly use the amplitude error model, where imperfect
rotations take the form $M(\theta, \phi) = R(\theta(1+\ep{A}),\phi)$.

We now show how a Solovay-Kitaev sequence with first-order broadband
properties may be derived.  Although the method presented in section
\ref{method} may be applied, here a more direct technique is used.
Our strategy is to construct a pulse sequence entirely out of
imperfect rotations $M(\theta, \phi) = R(\theta(1+\ep{A}),\phi) =
\R(\theta \ep{A},\phi) R(\theta, \phi)$ using \eqn{eq:SK1-unaddressed}
as a template.  Explicitly, this is achieved with the matrix
manipulations
\begin{eqnarray*}
R(\theta,0) + \O(\ep{A}^2)&=& R(2\pi \ep{A},-\p{SK1})R(2\pi\ep{A},\p{SK1})
R(\theta\ep{A},0)R(\theta,0)\\
&=& R(2 \pi \ep{A}, -\p{SK1}) R(2 \pi, -\p{SK1}) R(2 \pi \ep{A}, \p{SK1})  R(2 \pi , \p{SK1}) R(\theta \ep{A}, 0) R(\theta,0),
\end{eqnarray*}
where we have right-multiplied \eqn{eq:SK1-unaddressed} by
$R(\theta,0)$ (after substituting $\ep{A}$ for $\ep{N}$) and inserted
the identity $R(2 \pi, \p{SK1}) = R(2\pi, -\p{SK1}) = -\Id$.  Then, by
combining rotations about the same axis, one obtains the result,
\begin{eqnarray}
  U((1+\ep{A})\u(t)) &=& R(2 \pi (1+\ep{A}), -\p{SK1}) R(2 \pi (1 + \ep{A}),
  \p{SK1}) R(\theta (1 + \ep{A}), 0) \nonumber \\
  M_\mathrm{SK1}(\theta,0) &=& M(2 \pi, -\p{SK1}) M(2 \pi, \p{SK1}) M(\theta,0)  = R(\theta,0) + \O(\ep{A}^2)
\end{eqnarray}
where again $\p{SK1} = \arccos(- \theta / 4\pi)$.  In the presence of
unknown amplitude errors, the sequence reduces the effect of the error
while applying the effective rotation $R(\theta,0)$.  Observe that in
terms of imperfect rotations this is the same SK1 sequence derived
earlier.  SK1 has both narrowband and broadband behavior, and the
sequence may correct both addressing and amplitude errors
simultaneously.  Such a sequence is called a \emph{passband} sequence.

\emph{Generalization to arbitrary gates in $\SU{2}$:} The sequence SK1
is designed to compensate single qubit rotations about the $\Hx$ axis.
If this sequence is to be useful in quantum computation, it must be
generalized so that any single qubit rotation may be corrected.  One
method involves transforming the sequence by a similarity
transformation of the pulse propagators. As an example, suppose we
require an SK1 sequence that performs the rotation $R(\theta,\phi) =
\exp(-\i \phi \Hz ) R(\theta,0) \exp(\i \phi \Hz)$ on the addressed
spins.  Similarity transformation of pulses under $\exp(-\i \phi \Hz
)$ represent a phase advance in the rotating frame.  It is evident
that the transformed sequence
\begin{eqnarray}
M_\mathrm{SK1}(\theta,\phi) = M( 2\pi, \phi-\p{SK1}) M( 2\pi ,
\phi+\p{SK1}) M(\theta, \phi)
\end{eqnarray}
performs the desired compensated rotation.  In this manner, a
compensating pulse sequence for any target operation $U_T \in \SU{2}$
may be constructed.  One may solve for the operation $\Upsilon$ that
performs the planar rotation $\Upsilon U_T \Upsilon^\dagger =
R(\theta,0)$, where $\theta = \hsnorm{ \log U_T } / \hsnorm{\Hx}$.
Then the transformed sequence $\Upsilon^\dagger
M_{\mathrm{SK1}}(\theta,0) \Upsilon$ performs a first-order
compensated $U_T$ operation.  Similarity transformations of pulse
sequence propagators can be extremely useful, and are frequently
applied in composite pulse sequences.

Alternative methods exist for generating an arbitrary compensated
rotations.  Recall from section \ref{euler} that any operation $U_T
\in \SU{2}$ may be expressed in terms of a Euler decomposition $U_T =
R(\alpha_3, 0)R(\alpha_2, \pi/2)R(\alpha_1, 0)$. An experimentalist
may apply $U_T$ by implementing each of the Euler rotations in
sequence, however in the presence of an unknown systematic error each
of the applied rotations is imperfect and the fidelity of the applied
gate is reduced.  The error may be compensated by replacing each
imperfect pulse with a compensating pulse sequence.  For example, the
sequence
\begin{eqnarray}
    M_{\mathrm{SK1}}(\alpha_3,0) M_{\mathrm{SK1}}(\alpha_2,\pi/2) M_{\mathrm{SK1}}(\alpha_1,0) = U_T + \O(\ep{A}^2)
\end{eqnarray}
compensates amplitude errors to first-order by implementing SK1
sequences for each of the rotations in an Euler decomposition for
$U_T$.  This construction is an example of pulse sequence
concatenation.  By concatenating two independent pulse sequences, it
is sometimes possible to produce a sequence with properties inherited
from each parent sequence.


\subsubsection{Arbitrarily accurate SK sequences}
\label{SK-method}
In this section, we discuss the Solovay-Kitaev method for constructing
arbitrarily accurate composite pulse sequences, which may be used to
systematically improve the performance of an initial seed sequence.
The Lie algebraic picture is particularly helpful in the description
of the algorithm.  The method is quite general, and can be used on
sequences other than SK1.

Suppose that we have an $n$th-order compensating pulse
sequence, here denoted as $W_n$.  The problem we consider is the
identification of a unitary operator $A_{n+1}$ such that $W_{n+1} =
A_{n+1} W_{n} = U_T + \O(\epsilon^{n+2})$, where $W_{n+1}$ is an
$(n+1)$th-order sequence.  Assume for now that such an
operator exists, and consider that in the presence of systematic
errors, the application of the correction gate $A_{n+1}$ is imperfect.
However, if it is possible to implement a compensating pulse sequence
$B_{n+1}$, which is an $\O(\epsilon^{n+2})$ approximation of $ A_{n+1}$,
then it is still possible to construct an $(n+1)$th-order
sequence, .
\begin{eqnarray}
W_{n+1} = B_{n+1} W_{n} = U_T + \O(\epsilon^{n+2}).
\end{eqnarray}
We may then continue constructing pulse sequences of increasing
accuracy in this fashion if there exists a family of operators
$\{A_{n+1}, A_{n+2}, A_{n+3}, \cdots A_{m}\}$ and a corresponding family of
pulse sequences $\{B_{n+1}, B_{n+2}, B_{n+3}, \cdots B_{m}\}$ which
implement the operators to the required accuracy.  This immediately
suggests an inductive construction for the sequence $W_m$, 
\begin{eqnarray}
W_{m} = B_{m} \, B_{m-1} \cdots B_{n+3} \, B_{n+2} \, B_{n+1} \, W_n = U_T + \O(\epsilon^{m+1}).
\end{eqnarray}
This is the basis of the Solovay-Kitaev method \cite{Brown2004}.  To
apply the method, we must first have a means of calculating the
correction $A_{n+1}$, and second we must find a compensating sequence
$B_{n+1}$ robust to the systematic error model considered.

We turn our attention to the calculation of the correction terms
$A_{n+1}$.  It is convenient to decompose the pulse sequence
propagator using an interaction frame as $W_n = U_T U^I(\epsilon
\du(t))$, where $U_T = U(\u(t))$ is the target gate.  In the spirit of
\eqn{eq:magnus-cancellation}, a Magnus expansion for $U^I(\epsilon
\du(t))$ may be used.  Observe that $W_n = U_T \exp( \epsilon^{n+1}
\Omega_{n+1} ) + \O(\epsilon^{n+2})$.  Then letting $A_{n+1} = U_T
\exp( - \epsilon^{n+1} \Omega_{n+1} ) U^\dagger_T$ it may be verified
using the BCH formula that $A_{n+1} W_n = U_T + \O(\epsilon^{n+2})$.
This result may also be interpreted in terms of vector displacements
on the Lie algebra.  In the interaction frame, $\epsilon^{n+1}
\Omega_{n+1}$ may be interpreted as a vector in $\su{2}$.  Similarly,
the infinitesimal rotation $A_{n+1}$ corresponds to the vector
$-\epsilon^{n+1} \Omega_{n+1}$, equal in magnitude and opposite in
orientation.  The first-order term of the BCH series corresponds to
vector addition on the Lie algebra, and the $\O(\epsilon^{n+1})$ terms
cancel.

What remains is to develop a compensating pulse sequence that
implements $A_{n+1}$ to the required accuracy under a given error
model.  Let us define $P_{j z}(\alpha) = \exp(-\i \alpha \epsilon^j
\Hz) + \O(\epsilon^{j+1})$ and also the rotated analogues $P_{j
  x}(\alpha) = \exp(-\i \alpha \epsilon^j \Hx) + \O(\epsilon^{j+1})$
and $P_{j y}(\alpha) = \exp(-\i \alpha \epsilon^j \Hy) +
\O(\epsilon^{j+1})$. Frequently if one such $P_j$ may be produced,
then often the remaining two may be produced by similarity
transformation of the propagators or by using an Euler decomposition.
At this point we use \eqn{SK}, used in the proof for the
Solovay-Kitaev theorem, to construct relation
\begin{eqnarray}
P_{kx}(-\alpha) P_{\ell y}(-\beta) P_{kx}(\alpha) P_{\ell y}(\beta) &=& P_{jz}(\alpha \beta), \qquad k + \ell = j.
\label{eq:recursive-Pj}
\end{eqnarray}
Continuing in this manner, each of the $P_j$'s may be recursively
decomposed into a product of first-order propagators $P_{1 k}(\alpha)
= \exp(-\i \alpha \epsilon H_\mathrm{k} ) + \O(\epsilon^2)$.  Our
strategy is to use \eqn{eq:recursive-Pj} to implement $P_{(n+1)
    z}(\xi)$ where $\xi = \hsnorm{\Omega_{n+1}}/\hsnorm{\Hz}$.  On
the Lie algebra, this operation is represented by vector of length
$\epsilon^{n+1} \hsnorm{\Omega_{n+1}}$ oriented along the $-\i \Hz$
axis.  Let $\Upsilon$ be the rotation which performs $\Upsilon
\Omega_{n+1} \Upsilon^\dagger = \i \xi \Hz$ (i.e., the operator which
rotates $\Omega_{n+1}$ onto the $\i \Hz$ axis).  Then, by similarity
transformation under $U_T \Upsilon^\dagger$
\begin{eqnarray}
U_T \Upsilon^\dagger P_{(n+1) z}(\xi) \Upsilon U^\dagger_T = U_T \exp( -\epsilon^{n+1} \Omega_{n+1} ) U^\dagger_T + \O(\epsilon^{n+2}) = B_{n+1} ,
\label{eq:Bn}
\end{eqnarray}
the sequence $P_{(n+1) z}(\xi)$ may be transformed into precisely the
required correction sequence needed for the Solovay-Kitaev method.
There are two approaches for applying the transformation by $U_T
\Upsilon^\dagger$: we may either calculate the transformed analogues
of each of the pulses in $P_{(n+1)z}(\xi)$ and then apply the
transformed pulses, or we may, when possible, directly include the
transformation pulses (or an estimate for them) as physically applied
pulses in the sequence.  The second approach is only viable when it is
possible to generate accurate inverse operations $\Upsilon
U^\dagger_T$ \cite{Alway2007}.

Our construction will be complete once we have a method for generating
the simple ``pure-error'' propagator $P_{1x}(\alpha)$.  In many cases
it is sufficient to only consider $P_{1x}(\alpha)$ since the other
propagators $P_{1y}(\alpha)$ and $P_{1z}(\alpha)$ are related by a
similarity transformation.  In general, the method will depend on the
error model under consideration; here we explicitly show how to
construct this term in the amplitude and pulse-length error models and
also in the addressing error model.

\begin{figure}
\begin{center}
\includegraphics{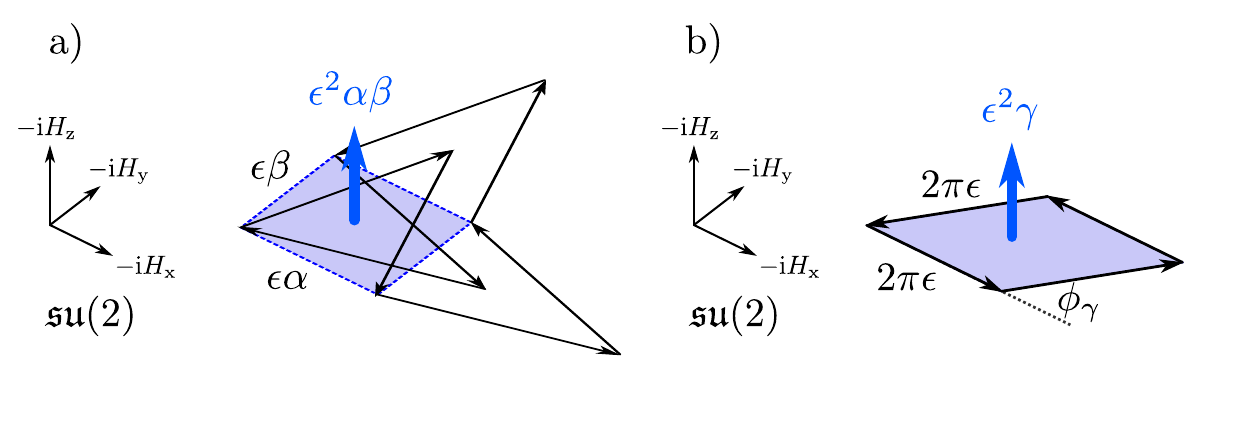}
\end{center}
\caption{Generation of the pure error term in the Solovay-Kitaev
    method. a) From \eqn{eq:recursive-Pj}, $P_{2z}(\alpha \beta)$ may
    be produced by the sequence $S_x(-\alpha) S_y(-\beta) S_x(\alpha)
    S_y(\beta)$.  On the Lie algebra the sequence corresponds to a
    closed rectangular path with an enclosed area of $\epsilon^2
    \alpha \beta$.  b) Alternatively, the rhombus construction may be
    used to generate $P_{2z}(\alpha)$ using four
    pulses. \label{fig:SK-method}}
\end{figure}

\emph{Addressing errors:} We wish to perform the evolution
$P_{1x}(\alpha)$ using a product of imperfect square-pulse propagators
$M(\theta,\phi)$.  Recall that in this model rotations on the
addressed qubit are error free $M(\theta,\phi) = R(\theta,\phi)$,
whereas on the unaddressed qubit the applied unitary depends on the
systematic addressing error $M(\theta,\phi) = R(\theta \ep{N}, \phi)$.
Similarly, the sequence implementing $P_{1x}(\alpha)$ must resolve to
the identity on the addressed qubit, while on the unaddressed qubit
$\exp(-\i \alpha \ep{N} \Hx) + \O(\ep{N}^2)$ is applied.  This
behavior may be achieved by using a pulse sequence to implement
$P_{1x}$.  Let
\begin{eqnarray}
S_x(\alpha) = M(2 \pi a,-\phi_\alpha) M(2 \pi a, \phi_\alpha),
\label{eq:Sx}
\end{eqnarray}
where $a = \lceil |\alpha| / 4 \pi \rceil$ is a integer number of
$2\pi$ rotations and $\phi_\alpha = \arccos( \alpha / 4 \pi a )$.  On the
addressed qubit, $S_x(\alpha)$ resolves to the identity, whereas for
the unaddressed spins,
\begin{eqnarray*}
S_x(\alpha) = M(2\pi a, -\phi_\alpha) M(2\pi a, \phi_\alpha) = R(2\pi a \ep{N}, -\phi_\alpha) R(2\pi a \ep{N}, \phi_\alpha) = P_{1x}(\alpha),
\label{eq:Sx-sequence}
\end{eqnarray*}
is applied.  In the Lie algebraic picture, $S_x(\alpha)$ is composed
of two vectors constructed so that their vector sum is $-\i \alpha
\ep{N} \Hx$. Similarly, $S_y({\beta})= M(2\pi b,\pi/2 -\phi_\beta)
M(2\pi b, \pi/2+\phi_\beta) = P_{1y}(\beta)$.  From these basic
sequences, we may construct $P_{2z}(\alpha \beta)$ using by using the
balanced group commutator \eqn{eq:recursive-Pj}.  In figure
\ref{fig:SK-method}a we plot the sequence $P_{2z}(\alpha \beta)$ as a
vector path on the Lie algebra.  The sequence encloses a signed area
$\ep{N}^2 \alpha \beta$, which is denoted by the shaded rectangular
figure.  By tuning the rotation angles $\alpha$ and $\beta$, one may
generate a term which encloses any desired area, thus allowing the
synthesis of an arbitrary pure-error term.

At this point, we discuss a subtle feature of the addressing error
model which at first appears to complicate the application of the SK
method.  Observe that on the unaddressed spin, the imperfect
propagators may only apply small rotations (i.e. rotations by angles
$\theta \ep{N}$).  If we restrict ourselves to sequences composed of
resonant square pulses, then the term proportional to $P_{1z}(\alpha)$
may not be produced; we may not prepare such a term by similarity
transformation (e.g. $R(\pi/2,0) S_y(\alpha) R^\dagger(\pi/2,0)$)
since such an operation would either require a large rotation or if
instead the transformation was carried out on the individual sequence
propagators, the rotation axes would be lifted out of the $\Hx$-$\Hy$
plane.  Similar arguments show that the Euler decomposition is also
unavailable.

Fortunately, this restriction is not as serious as it first appears;
the SK method may be used provided that the sequence terms are chosen
with care.  We are ultimately saved by the orientation of the error
terms in the Lie algebra.  Using the BCH formula it is straightforward
to show that for sequences composed of resonant pulses, the even-order
error terms are always aligned along the $-\i \Hz$ axis, whereas the
odd-order terms are confined to $\Hx -\Hy$ plane.  Likewise, 
using only $S_{x}(\alpha)$ and
$S_{y}(\beta)$, it is possible to generate correction terms that
follow the same pattern.  As a consequence, in this case it is
possible to generate the correction terms $U_T \exp(-\ep{N}^{n+1}
\Omega_{n+1}) U_T^\dagger$ by carefully choosing the rotation angles
and phases in the correction sequence $B_{n+1}$.

As a instructive example, we shall derive a second-order passband
sequence using the Solovay-Kitaev method.  We begin by calculating
the Magnus expansion for the seed sequence $M_{\mathrm{SK1}}(\theta,0)
= U_T \exp( \ep{N}^2 \Omega_2 + \ep{N}^3 \Omega_3 + \cdots )$ where
the target operation $U_T = \Id$ for the unaddressed qubit. To cancel
the second order term, we simply need to apply the inverse of $\exp(
\ep{N}^2\Omega_2) = \exp(-\i 2\pi^2 \ep{N}^2 \sin(2\p{SK1})\Hz)$. The planar 
rotation $\Upsilon = \Id$,
since $B_2 = P_{2z}(-2 \pi^2 \sin(2\p{SK1}))$ is already oriented in the
correct direction.  One possible choice for $B_2$ is the sequence
\begin{eqnarray}
B_2 =S_x(-2\pi \cos \p{SK1}) S_y(2\pi \sin \p{SK1}) S_x(2 \pi \cos \p{SK1}) S_y(-2 \pi \sin \p{SK1}). 
\end{eqnarray}  
The sequence $M_\mathrm{SK2}(\theta,0) = B_2 M_\mathrm{SK1}(\theta,0)
= U_T + \O(\ep{N}^3)$ corrects addressing errors to second order.  We
denote an $n$th order compensating sequence produced by the
Solovay-Kitaev method using SK1 as an initial seed as SK$n$; here we
have produced an SK2 sequence.

More efficient constructions for the correction sequence $B_2$
exist. Observe that we may directly create the pure error term
$P_{2z}(\gamma)$ by using four pulses in the balanced group commutator
arrangement $ P_{2z}(\gamma) = M^\dagger (2\pi c, \phi_\gamma )
M^\dagger(2\pi c , \phi'_\gamma ) M(2\pi c, \phi_\gamma) M(2\pi c,
\phi'_\gamma)$, where $c = \lceil |\gamma| / 4\pi \rceil$ and the
phases are chosen to be $\phi_\gamma = \arcsin( \gamma / 4 \pi^2 c^2
)$ and $\phi'_\gamma = (1-\mathrm{sign}\gamma) \pi/2$.  The phase
$\phi'_\gamma$ is only necessary to ensure that the construction also 
works for negative $\gamma$.
We call this arrangement the rhombus construction.  In figure
\ref{fig:SK-method}b we plot this sequence as a vector path on
$\su{2}$.  In this construction the magnitude of the error term is
tuned by adjusting the phase $\phi_\gamma$, i.e. adjusting the area
enclosed by the rhomboidal path of the sequence in the Lie algebra.
The rhombus construction has the advantage of requiring half as many
pulses as the standard method.  We will now use this construction to
produce an alternative form for SK2.  Let
\begin{eqnarray}
B_2' = M(2\pi,\phi_\gamma + \pi) M(2\pi, 0) M(2\pi,\phi_\gamma) M(2\pi, \pi) 
\end{eqnarray}
where $\phi_\gamma = \arcsin( \sin(2 \p{SK1}) / 2 )$.  Then
$M_{\mathrm{SK2}}'(\theta,0) = B_2' M_{\mathrm{SK1}}(\theta,0) = U_T + \O(\ep{N}^3)$.

\emph{Amplitude / pulse-length errors:} We now turn our attention to
the compensation of amplitude and pulse-length errors.  When
considering these error models, the imperfect propagators of the form
$M(\theta,\phi) = R(\theta (1 + \ep{A}),\phi)$.  In this case we may
also construct $P_{1x}$ using \eqn{eq:Sx} since the imperfect $2\pi$
rotations reduce to $M(2 \pi a, \phi) = R(2 \pi a \ep{A}, \phi)$ and
$S_x(\alpha) = P_{1x}(\alpha)$.  As a consequence, if the initial seed
sequence $W_n$ is a passband sequence then $W_{n+1}$ is also a
passband sequence.  We note however, that in this error model one has
more flexibility in the synthesis of the correction sequences
$B_{n+1}$; since now the imperfect propagators apply large rotations,
then we may perform similarity transformations of sequences by
directly implementing the required pulses.  Notably, we may use
imperfect propagators of order $\O(\ep{A}^n)$ to apply a desired
transformation at a cost of an error of order $\O(\ep{A}^{n+1})$, for
example $M(\theta,\phi) P_{jz} M^\dagger(\theta,\phi) = R(\theta,\phi)
P_{jz} R^\dagger(\theta,\phi) + \O(\ep{A}^2)$.

The SK method may be used to calculate higher order compensation
sequences for the amplitude error model.  We begin by calculating the
relevant Magnus expansion for the seed sequence
$M_\mathrm{SK1}(\theta,0) = M(2\pi,-\p{SK1}) M(2\pi,\p{SK1}) M(\theta,0) =
U_T \exp( \ep{A}^2 \Omega_2 + \ep{A}^3 \Omega_3 + \cdots )$ where now
in the amplitude error model the target operation $U_T=R(\theta,0)$.
To cancel the second order term we must apply the inverse of $U_T
\exp(\ep{A}^2\Omega_2) U_T^\dagger$ = $\exp(-\i2\pi^2 \ep{A}^2
\sin(2\p{SK1})\Hz)$.  This is precisely the same term that arose
previously for addressing errors and the error may be compensated the
same way.  In fact every $n$th order SK sequence is passband and works
for both error models.


\subsection{Wimperis / Trotter-Suzuki sequences}
\label{wimperis}
In this section we study a second family of fully compensating pulse
sequences, first discovered and applied by Wimperis
\cite{Wimperis1994}, which may be used to to correct pulse length,
amplitude, and addressing errors to second order. 
These sequences have been remarkably successful and have found extensive use in 
NMR and quantum information \cite{Morton2005, Xiao2006, Beavan2009}.
Furthermore, the Wimperis sequences may be
generalized to a family of arbitrarily high order sequences by
connecting them to Trotter-Suzuki formulas \cite{Brown2004}.  An
analogous composite sequence composed of rotations in $\SU{4}$ may be
used to correct two-qubit operations \cite{Jones2003, Tomita2010}.  We study
the two-qubit case in section \ref{two-qubit}.

\emph{Narrowband behavior:} We begin
with the problem of identifying narrowband sequences which correct
addressing errors.  In the addressing error model operations on
addressed qubits are error free, whereas on the unaddressed qubits the
imperfect propagators takes the form $V(\u(t)) = U(\ep{N} \u(t))$ and
$U_T = \Id$. Further, we shall constrain ourselves to sequences
composed of resonant square-pulse propagators.  Specifically, we
search for arrangements of four pulses that eliminate both the first
and second-order Magnus expansion term for the imperfect
propagator.  

Before explicitly describing the construction of the Wimperis
sequences, we digress shortly to point out a certain symmetry property
which may be used to ensure that $\ep{N}^2 \Omega_2 = 0$.  Consider
the group product of propagators of the form,
\begin{eqnarray}
  U(\ep{N} \u(t)) = U_2 U_3 U_2 U_1, \qquad U_k = \exp(-\i \ep{N} t_k\u_k \cdot \vec{H}),
\label{eq:symmetry}
\end{eqnarray}
with the added condition that the first-order expansion term for
$U(\ep{N}\u(t))$ has been already eliminated, i.e. $\ep{N} \Omega_1 =
-\i \ep{N} (t_3\u_3 + 2 t_2\u_2 + t_1\u_1) \cdot \vec{H} = 0$.  In this
arrangement, the second and fourth propagators are identical;  this pulse
symmetry along with $ t_1\u_1+2t_2\u_2+t_3\u_3= 0$ eliminates the
second-order term,
\begin{eqnarray}
\ep{N}^2 \Omega_2 &=& \frac{\ep{N}^2}{2} \sum_{i=1}^4 \sum_{j=1}^i [ -\i t_i \u_i \cdot \vec{H}, -\i t_j \u_j \cdot \vec{H} ]=0. 
\label{eq:symmetry-cancel}
\end{eqnarray}
As a consequence, $U(\ep{N} \u(t)) = \Id + \O(\ep{N}^3)$. 
Alternatively, one may regard the
product $T = U_2U_3U_2$ as a second-order symmetric Trotter-Suzuki
formula for the inverse operation $U_1^\dagger$ \cite{Suzuki1992}
which approximately cancels the undesired rotation $U_1$. Considering
the control fields applied during the application of the corrector
sequence $T$, the applied control Hamiltonian is symmetric with
respect to time inversion.  By a well known theorem, all even-order
expansion terms produced by a time-symmetric Hamiltonian cancel
\cite{Burum1981, Levitt2008} (i.e. $\ep{N}^{2j} \Omega_{2j} = 0$,
for all positive integers $j$).  Thus by implementing symmetric
corrector sequences $T_{2j}$, it is sufficient to only consider the
cancellation of the remaining odd-order error terms.  In section
\ref{arbitrarily-accurate-TS} we will inductively develop a series of
corrector sequences of increasing accuracy based on symmetric
Trotter-Suzuki formulas \cite{Suzuki1992, Brown2004}.

\begin{figure}
\begin{center}
\includegraphics{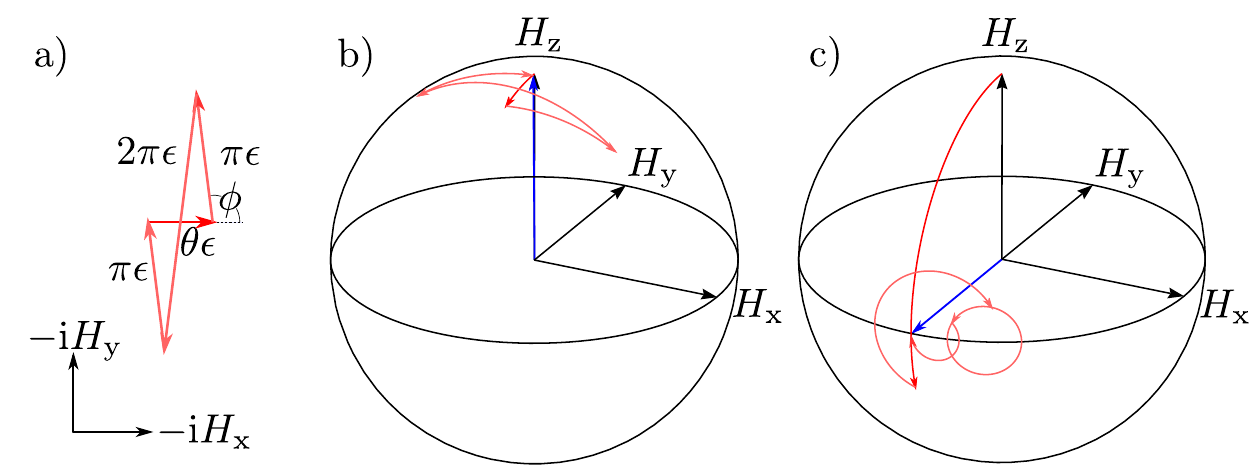}
\end{center}
\caption{a) Vector path followed by N2 on the Lie algebra. b)
    Trajectory of an unaddressed spin during an N2 sequence, using
    imperfect rotations of the form $M(\theta,\phi) =
    R(\theta\ep{A},\phi)$.  c) B2 correcting an amplitude error, using
    imperfect rotations of the form $M(\theta, \phi) =
    R(\theta(1+\ep{A}),\phi)$.  In these plots $\ep{N} = \ep{A} =
    0.2$.
   \label{figure-wimperis}}
\end{figure}

The cancellation of the second-order term may also be inferred from
geometric considerations on the Lie algebra.  To be concrete, consider
a sequence of the form \eqn{eq:symmetry} where
\begin{eqnarray}
\ep{N} t_{1}\u_1\cdot\vec{H} &=& \theta \ep{N} \Hx \nonumber \\
\ep{N} t_{2}\u_2\cdot\vec{H} &=& \pi \ep{N}( \cos\p{N2} \Hx + \sin \p{N2} \Hy ) \nonumber \\
\ep{N} t_{3}\u_3\cdot\vec{H} &=& 2 \pi \ep{N}( \cos\p{N2} \Hx - \sin \p{N2} \Hy ),
\label{eq:nb1-vectors}
\end{eqnarray}
and the phase is $\p{N2} = \arccos(-\theta/4\pi)$ is chosen so that $
t_1\u_1+2t_2\u_2+t_3\u_3= 0$, i.e., the vectors form a closed path on
the dynamical Lie algebra.  Figure \ref{figure-wimperis}a is a diagram
of these vectors on $\su{2}$.  In $\su{2}$, the Lie bracket is
equivalent to the vector cross product \cite{Gilmore}, therefore we
may interpret $\hsnorm{\ep{N}^2 \Omega_2}$ as the signed area enclosed
by the vector path on the Lie algebra.  We note that the sequence
under study encloses two regions of equal area and opposite sign,
ensuring that the second order term is eliminated.

With this insight, the construction of a second-order narrowband
sequence is straightforward.  The vectors \eqn{eq:nb1-vectors}
correspond to the sequence  
\begin{eqnarray}
  M_\mathrm{N2}(\theta,0) = M(\pi,\p{N2}) M(2\pi,-\p{N2}) M(\pi,\p{N2}) M(\theta, 0).
\label{eq:N2}
\end{eqnarray}
Wimperis refers to this sequence as NB1, and indeed, this is the
established name in the literature.  In the current article, we label
this sequence N2 in anticipation of the generalization of this form to
N2$j$, which compensates addressing errors to $\O(2j)$.  We use this
language to avoid confusion with other established sequences, namely
NB2, NB3, etc \cite{Wimperis1994}.  The N2 sequence may be used to
compensate addressing errors.  For unaddressed qubits $M(\theta,\phi)
= R(\theta \ep{N},\phi)$ and thus from \eqn{eq:symmetry-cancel} and
\eqn{eq:nb1-vectors} it follows that $M_{\mathrm{N2}}(\theta,0) = \Id
+ \O(\ep{N}^3)$.  Thus on unaddressed qubits the sequence performs the
identity operation up to second-order.  Furthermore, for addressed
qubits $M(\theta,\phi) = R(\theta,\phi)$ and therefore
$M_{\mathrm{N2}}(\theta,0) = R(\theta,0)$.  As a result, when the
sequence $M_\mathrm{N2}(\theta,0)$ is used in the place of the
imperfect operation $M(\theta,0)$, the discrimination between
addressed and unaddressed spins is enhanced.  In figure
\ref{figure-wimperis}b we plot the magnetization trajectory for an
unaddressed qubit under an N2 sequence.

\emph{Broadband behavior:} As previously discussed, broadband
sequences are best suited for correcting amplitude or pulse-length
errors.  In the following we will explicitly consider the amplitude
error model, where $M(\theta,\phi) = R(\theta(1 + \ep{A}),\phi) =
R(\theta,\phi) R(\theta \ep{A},\phi)$.  Although a pulse sequence may
be studied by considering the interaction frame propagator as
described in section \ref{method}, the method originally used by
Wimperis is simpler.  Wimperis' insight was that for lowest orders the
toggled frame could be derived geometrically, using the relation
$R(\theta,-\phi)= R(\pi,\phi) R(\theta,3\phi) R(\pi,\phi+\pi)$.
Consider the application of the target gate $U_T = R(\theta,0)$ using
the sequence
\begin{eqnarray}
M_\mathrm{B2}(\theta,0) = M(\pi,\p{B2}) M(2\pi,3\p{B2}) M(\pi,\p{B2}) M(\theta, 0)
\end{eqnarray}
where $\p{B2} = \arccos(-\theta/4\pi)$.  This is the Wimperis
broadband sequence, traditionally called BB1 but here denoted as B2.
When rewritten in terms of proper rotations one obtains,
\begin{eqnarray*}
  M_\mathrm{B2}(\theta,0) &=&  R(\pi \ep{A}, \p{B2}) \Big( R(\pi,\p{B2}) R(2\pi \ep{A}, \p{B2}) R^\dagger(\pi, \p{B2}) \Big) R(\pi \ep{A}, \p{B2} ) \R(\theta \ep{A} , 0) R(\theta, 0) \nonumber \\
&=& \Big[ R(\pi \ep{A}, \p{B2}) R(2\pi \ep{A}, -\p{B2}) R(\pi \ep{A}, \p{B2}) R(\theta \ep{A}, 0) \Big] R(\theta,0)
\end{eqnarray*}
where the identity $R(2\pi,3\p{B2}) = -\Id = R(\pi,\p{B2} + \pi)
R(\pi,\p{B2} + \pi)$ was used.  Let $Q$ denote the quantity enclosed in
square brackets, so that we may write $M_{\mathrm{B2}}(\theta,0) = Q
U_T$.  Observe that $Q$ is precisely the form considered previously in
N2.  From our previous result we may conclude $Q = \Id + \O(\ep{A}^3)$
and $M_{\mathrm{B2}}(\theta,0) = R(\theta,0) + \O(\ep{A}^3)$.  As a
result, when $M_\mathrm{B2}(\theta,0)$ is used in the place of the
imperfect operation $M(\theta,0)$ the effect of the systematic
amplitude error is reduced.  Note that errors for the B2 and N2
sequences follow equivalent paths on the Lie algebra in their
respective interaction frames.  In figure \ref{figure-wimperis}c we
plot the magnetization trajectory for a qubit under a B2 sequence with
amplitude errors.

\emph{Passband behavior:} In some cases, it is convenient to have a
passband pulse sequence that corrects for both addressing errors and
for amplitude errors, as we saw with the Solovay-Kitaev sequences.
The passband Wimperis sequence P2 is simply two Solovay-Kitaev
correction sequences in a row where the order of pulses is switched,
$M_\mathrm{P2}(\theta,0) = M(2\pi,\p{P2}) M(2\pi,-\p{P2})
M(2\pi,-\p{P2}) M(2\pi,\p{P2}) M(\theta,0)$, with $\p{P2}=
\arccos(-\theta/8\pi)$.  The switching of pulse order naturally
removes the second order error term.  One may verify that this
sequence works for both addressing and amplitude errors.


\subsubsection{Arbitrarily accurate Trotter-Suzuki sequences}
\label{arbitrarily-accurate-TS}
The Wimperis sequences rely on a certain symmetrical ordering of pulse
sequence propagators to ensure that even order Magnus expansion terms
are eliminated.  We may further improve the performance of these
sequences by taking advantage of additional symmetries that cancel
higher-order terms.  In the following, we shall show how by using
symmetric Trotter-Suzuki formulas \cite{Trotter1958, Suzuki1993,
  Suzuki1992} a family of arbitrarily accurate composite pulse
sequences may be constructed \cite{Brown2004, Alway2007}.  The Lie
algebraic picture along with the Magnus and BCH series will be
important tools in this process.

\emph{Symmetrized Suzuki formulas:} Before discussing the particular form of
these sequences, it is helpful to briefly mention a few important results
regarding symmetric products of time-independent propagators
\cite{Suzuki1992}.  Given a series of skew-Hermitian time-independent
Hamiltonians $\{ \tH_1, \tH_2, \dots , \tH_m \}$ such that $\sum_i^m
\tH_i = \tH_T$, the BCH expansion tells us
\begin{eqnarray}
\prod_{i=1}^{m} \exp(\lambda \tH_i) = \exp \left( \lambda \tH_T + \sum_{n=2}^\infty \lambda^n \Omega_n \right),
\end{eqnarray}
where $\lambda$ is a real parameter and the expansion terms $\Omega_n$
depend on the specific ordering of the sequence.  If we choose to
apply the operators in a time-symmetric manner such that $\exp(\lambda
\tH_i ) = \exp( \lambda \tH_{m+1-i} )$, then the symmetry of the pulse
removes all even-order terms.  For symmetric products, we have
\begin{eqnarray}
\prod_{symmetric} \exp(\lambda \tH_i) = \exp \left( \lambda \tH_T + \sum_{j=2}^\infty \lambda^{2j-1} \Omega_{2j-1} \right). 
\end{eqnarray}

An important observation concerning the elimination of the remaining
odd terms was made by Suzuki \cite{Suzuki1992}.  Provided that $\tH_i
= p_i \tH_T + (p_i)^{2j-1}\tH_B$, where the coefficients $p_i$ are
real numbers, there exist certain choices of coefficients such that
$\sum_i^m \tH_i = \tH_T$.  This requires that $\sum_i^m p_i=1$ and
$\sum_i^m (p_i)^{2j-1}=0$.  The situation is considerably simplified
if we restrict ourselves to sequences composed of just two kinds of
propagators, $U_1 = \exp(\tH_1)$ and $U_2 = \exp(\tH_2)$.  In this case 
the previous expression simplifies to $n_1 p_1
+ n_2 p_2 = 1$ and $n_1 p_1^{2j-1} +n_2 p_2^{2j-1} = 0$, where the
integers $n_1$ and $n_2$ are the number of $U_1$ and $U_2$ pulses
required to produce a sequence that is independent of $H_B$ up to
$\O(p^{2j+1})$.  We solve for a set of coefficients by setting
$p_2 = -2p_1$ and $n_2 = 1$, thus yielding $n_1 = 2^{2j-1}$ and $p_1=1/(2^{2j-1}-2)$.

Combining these observations, if $W_{2j-2}(p)$ is a $(2j-2)$th
approximation of $\exp( p \tH_T )$ and $\Omega_{2j-1}=p^{2j-1} \tH_B$
where $\tH_B$ is independent of $p$, then we can construct $W_{2j}(1)$
from the lower approximations, $W_{2j}(1)= (W_{2j-2}(p))^{2^{2j-2}}
W_{2j-2}(-2p)(W_{2j-2}(p))^{2^{2j-2}}$. 
As a result, Suzuki formulas provide a path of producing higher-order
sequences from a symmetric combination of lower-order sequences.
Notice that the symmetric decomposition is used to keep the even-order
terms zero.  
 
\emph{Passband behavior:} In the following we will seek to generalize
the second-order passband sequence P2 to an arbitrarily accurate
passband sequence P$2j$.  Our goal is to develop a correction sequence
$T_{2j}(k,\phi) = \exp( \i \theta \ep{A} \Hx ) + \O(\ep{A}^{2j + 1})$
that cancels the unwanted rotation of a $M(\theta,0)$ operation.  

When considering passband sequences, it is convenient to use rotation
angles that are integer multiples of 2$\pi$.  Let us define the
triangular motif,
\begin{eqnarray}
  T_1(k,\phi)&=& M(2 k \pi, -\phi) M(2 k \pi, \phi ) \nonumber \\
  &=&\exp(-\i 4k \pi \ep{A} \cos \phi \Hx -\i (2\pi k\ep{A})^2 \cos \phi \sin \phi \Hz ) + O(\ep{A}^3).
\end{eqnarray}
Observe the passband sequence SK1 may now be written as
$M_{\mathrm{SK}1}(\theta,0) = T_1(1,\p{SK1}) M(\theta,0)$ and where
the phase $\p{SK1}$ is as defined previously.
\begin{figure}
\begin{center}
\includegraphics{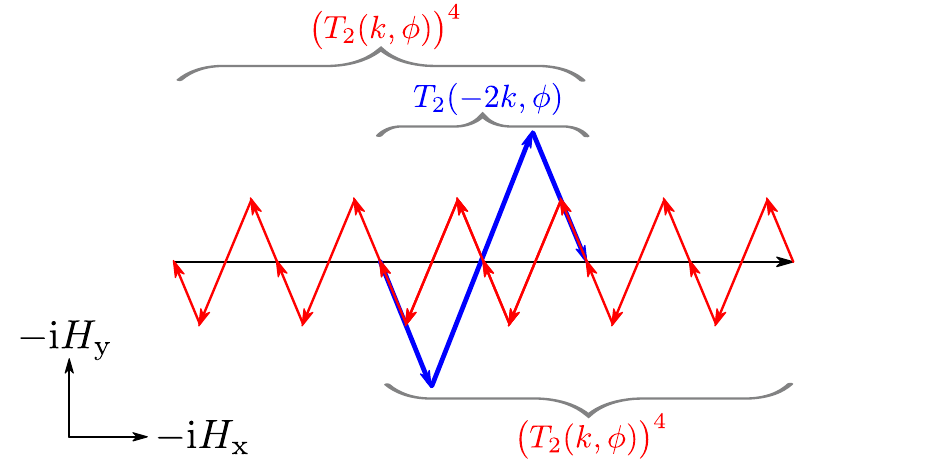}
\end{center}
\caption{Vector path followed by P4 on the Lie algebra.
   \label{figure-P4}}
\end{figure}
The remaining second-order term is odd with respect to $\phi$, and is
related to the vector cross product on the Lie algebra.  A
second-order sequence may be constructed by combining two
$T_1(k,\phi)$ terms so that the correction sequence is symmetric and
the cross product cancels. Let us define the symmetrized product
\begin{eqnarray}
T_2(k,\phi)&=&T_1(k,-\phi)T_1(k,\phi) \nonumber \\
&=& \exp( p \tH_T + p^3 \tH_B ) + \O(p^5),
\end{eqnarray}
where the length $p = - (8 k \pi/ \theta) \cos \phi $, the target
Hamiltonian $\tH_T = \i \theta \ep{A} \Hx$, and $\tH_B = \ep{A}^3
\Omega_3 / p^3$ is the remaining term we wish to cancel.  $\Omega_3$
is a function that depends on $k$ and $\phi$ such that $\tH_B$ is a
function of $\phi$ but not of the length scale $k$.  For fixed $\phi$
and variable $k$, this makes $H_B$ independent of $p$. Observe that
the passband sequence P2 may now be written as $M_{\mathrm{P}2}
(\theta, 0) = T_2(1,\p{P2}) M(\theta,0)$, where again $\p{P2} =
\arccos( - \theta / 8 \pi )$.

Our strategy is to use a Suzuki formula to construct higher-order
$T_{4}(k,\phi)$ (that is, $T_{2j}(k,\phi)$ for $j = 2$) using a
symmetric combination of $(n_1 = 8)$-many $\exp(\tH_1) = T_2(k,\phi)$
sequences and a single $\exp(\tH_2) = T_2(-2k,\phi)$ sequence,
yielding
\begin{eqnarray}   
T_4(k,\phi) = \Big(T_2(k,\phi) \Big)^4T_2(-2k,\phi) \Big(T_2(k,\phi) \Big)^4.
\end{eqnarray}
In order to produce to required correction term, the parameters
$(k,\phi)$ must be chosen such that $(n_1 - 2)p = 1$.  The
fourth-order passband sequence P4 is $M_{\mathrm{P4}}(\theta,0) =
T_4(1, \p{P4}) M(\theta,0)$, where $\p{P4} = \arccos( -\theta /48 \pi
)$.  In figure \ref{figure-P4} we plot the vector path followed by P4
on the Lie algebra.  This result may be further generalized.  To
produce $T_{2j}(k,\phi)$ requires $(n_1 = 2^{2j - 1})$-many $T_{2j -
    2}(k, \phi)$ sequences and a single $T_{2j - 2}(-2k, \phi)$
sequence in the symmetric ordering,
\begin{eqnarray}
T_{2j}(k,\phi)= \Big(T_{2j-2}(k,\phi) \Big)^{2^{2j-2}} T_{2j-2}(-2k,\phi) \Big( T_{2j-2}(k,\phi) \Big)^{2^{2j-2}}.
\end{eqnarray}
We then fix $\phi$ so that the first-order term cancels the unwanted
rotation, yielding
\begin{eqnarray}
\phi_{\mathrm{P2}j} = \arccos \left( - \frac{\theta}{2\pi f_j} \right)
\end{eqnarray}
where $f_j = (2^{2j - 1} -1) f_{j-1}$ and for the sequence P2$j$, $f_1
= 4$.  Then the $2j$th-order passband sequence P2$j$ is
$M_{\mathrm{P}2j} (\theta,0) = T_{2j} ( 1, \phi_{\mathrm{P}2j})
M(\theta,0)$.

\begin{figure}
\begin{center}
\includegraphics{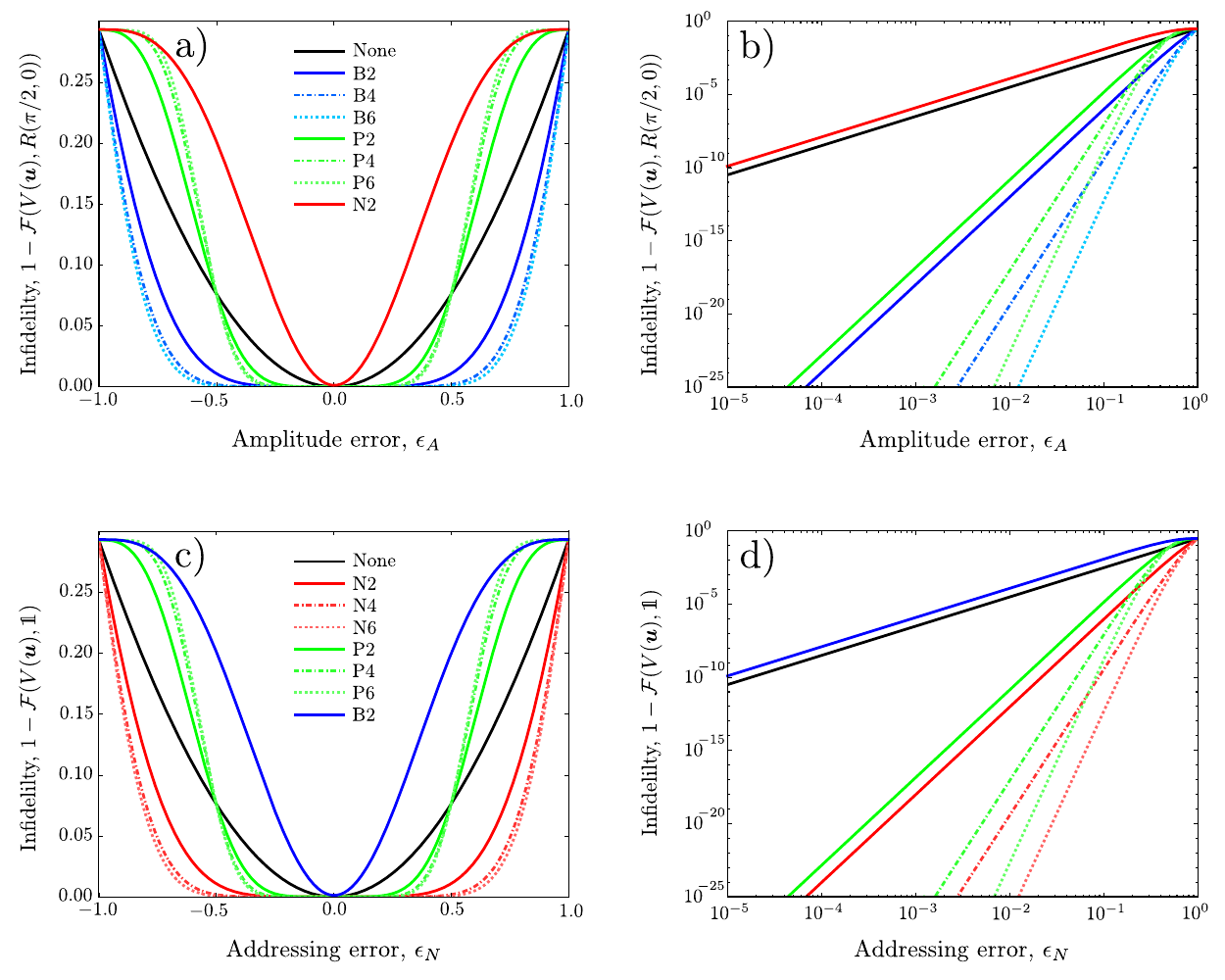}
\end{center}
\caption{Infidelity of the Trotter-Suzuki sequences B2$j$, P2$j$, and
    N2$j$.  In the amplitude error model (top row) $M(\theta,\phi) =
    M(\theta(1+\ep{A}),\phi)$ and $U_T = R(\pi/2,0)$.  In the
    addressing error model (bottom row) on the unaddressed qubits
    $M(\theta,\phi) = M(\theta \ep{N},\phi)$ and $U_T = \Id$, while on
    the addressed spins $R(\pi/2,0)$ is applied.  Each error model
    establishes a separate preferred interaction frame; when
    transformed into the appropriate pictures, the B2$j$ and N2$j$
    sequences are homologous.  The passband sequences P2$j$ can
    correct both amplitude and addressing errors at the cost of
    reduced efficacy.
  \label{figure-B2j-P2j}}
\end{figure}

The same method can be used to develop exclusively narrowband or
broadband sequences called N2$j$ and B2$j$ respectively
\cite{Brown2004}.  This requires redefining the bottom recursion layer
$T_2(k,\phi)$ to have either narrowband or broadband properties. For
N2$j$, $f_1 = 2$ and $T_2(k,\phi) = T_1(k/2,\phi)T_1(k/2,-\phi)$. For
B2$j$ $f_1 = 2$, but $T_2$ is slightly more complicated; when $k$ is
even $T_2(k,\phi) = T_1(k/2,-\phi)T_1(k/2,\phi)$ just like N2$j$,
however when $k$ is odd and $T_2(k,\phi) =
M(k\pi,\phi)M(k\pi,3\phi)M(k\pi,3\phi)$.  In figure
\ref{figure-B2j-P2j}, we compare several of the generalized
Trotter-Suzuki sequences to the ideal unitaries $U_T = R(\pi/2,0)$ in
the case of amplitude errors (top row) and $U_T = \Id$ in the case of
addressing errors on unaddressed qubits (bottom row).

\subsection{CORPSE}
So far, the sequences considered here have been designed to correct
systematic amplitude and addressing errors.  The correction of errors
arising from an inaccurate tuning of the control field are also of
practical interest.  The treatment of detuning errors is similar in
principle to the error models already considered, however in practice
the construction of compensating sequences is complicated by the
noncommutivity of the ideal Hamiltonian and the erroneous Hamiltonian
generated by the control distortion $[\u(t)\cdot \vec{H},
\du(t)\cdot\vec{H}] \neq 0$.

Fully compensating pulse sequences for detuning errors were originally
studied by Tycko \cite{Tycko1983} and later generalized by
Cummins and Jones into the humorously named ROTTEN (resonance offset
tailoring to enhance nuition) \cite{Cummins2001} and CORPSE
(compensating for off-resonance with a pulse sequence)
\cite{Cummins2000, Cummins2003} family of sequences.  Cummins and Jones
initially derived the sequence using quaternion algebra to represent
simple rotations, and optimized the angles to eliminate the first
order effects of the detuning error.  The CORPSE family of sequences
has found application in NMR \cite{Cummins2000} and SQUID
\cite{Collin2004} experiments.  Here we reexamine CORPSE using the
techniques outlined in section \ref{method}.


CORPSE is a sequence which performs a compensated rotation about the
$\Hx$ axis.  Following Cummins and Jones, the sequence is constructed
from three square pulses, which in the case of perfect resonance,
induce rotations about the $\Hx$, $-\Hx$, and $\Hx$ axes,
sequentially.  The pulse sequence is parametrized by the piecewise
constant control function
\begin{eqnarray}
u_\mathrm{x}(t) = \left \{ 
  \begin{array}{rlr}
      &\ux \qquad & 0 \leq t < t_1 \\
    - &\ux        & t_1 \leq t < t_2 \\
      &\ux        & t_2 \leq t \leq \tau
  \end{array}
\right .,
\label{eq:CORPSE-controls}
\end{eqnarray}
where $\ux$ is a constant $\Hx$ control amplitude and the $t_j$ are
times at which the field direction is switched.  This construction is
particularly amenable to analytic methods, as the ideal control
Hamiltonian $H(t) = \ux(t) \Hx$ commutes with itself at all times.  In
the absence of systematic detuning errors the control Hamiltonian
produces the following ideal unitary evolution,
\begin{eqnarray}
  U(\ux(t') \vec{\mathrm{x}}; t, 0 ) = \exp( -\i \vartheta(t) \Hx ) = R(\vartheta(t),0), \qquad \vartheta(t) = \int_0^t dt' \ux(t').
\label{eq:corpse-propagator}
\end{eqnarray}
Over the entire time interval $0 \leq t \leq \tau$, the pulse sequence
produces the gate $U_T = R(\theta,0)$, where $\theta = \vartheta(\tau)$.

\begin{figure}
\begin{center}
\includegraphics{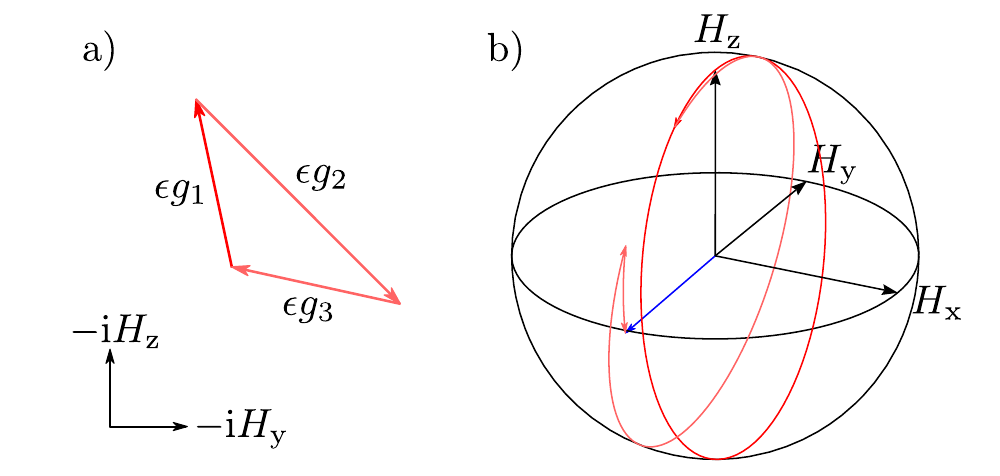}
\end{center}
\caption{a) Vector path followed by CORPSE on the Lie algebra, with
    the choice of parameters $n_1 = n_2 = n_3 = 0$.  Each vector $g_k$
    corresponds to a term in \eqn{eq:corpse-terms}. b) Trajectory of a
    spin under a CORPSE sequence for $U_T = R(\pi/2,0)$, with $n_1 =
    n_2 = 1$ and $n_3 = 0$ chosen to produce positive angles.  The
    sequence is constructed of imperfect rotations of the form
    $M(\theta_k,0) = \exp(-\i(\theta_k \Hx + \epsilon_D \theta_k
    \Hz))$, with $\epsilon_D = 0.2$. \label{fig:corpse}}
\end{figure}

In the presence of an unknown detuning error $\du = \ep{D}
\vec{\mathrm{z}}$ and the rotation axis is lifted in the direction of the
$\Hz$ axis on the Bloch sphere.  Recalling section \ref{method}, a
Magnus expansion may be used for the interaction frame propagator
$U^I(\ep{D} \vec{\mathrm{z}}; \tau, 0)$, which produces the evolution
generated by the systematic detuning error.  Combining
\eqn{eq:first-order} and \eqn{eq:corpse-propagator} the first-order
term is
\begin{eqnarray}
\ep{D} \Omega_1(\tau,0) = -\i \ep{D} \int_0^{\tau} dt \cos( \vartheta(t) ) \Hz + \sin( \vartheta(t) ) \Hy.
\label{eq:CORPSE-magnus}
\end{eqnarray}
Direct integration yields,
\begin{eqnarray}
{\textstyle \ep{D} \Omega_1(\tau,0) } \!\!\!\!&=&\!\!\!\!
{\textstyle - \frac{\i \ep{D}}{\ux} [ \sin(\theta_1) \Hz +(1 - \cos(\theta_1)) \Hy ] + \cdots} \nonumber \\
&\:& \!\!\!\!
{\textstyle + \frac{\i \ep{D}}{\ux}[ (\sin(\theta_1 - \theta_2) - \sin(\theta_1)) \Hz + (\cos(\theta_1)-\cos(\theta_1 - \theta_2)) \Hy ] + \cdots} \label{eq:corpse-terms}\\
&\:& \!\!\!\!
{\textstyle -\frac{\i \ep{D}}{\ux} [ (\sin(\theta_1 - \theta_2 + \theta_3) \!- \sin(\theta_1 - \theta_2)) \Hz \!+ (\cos(\theta_1-\theta_2) \!- \cos(\theta_1 - \theta_2 + \theta_3)) \Hy ],} \nonumber 
\end{eqnarray}
where $\theta_k = \vartheta(t_k) - \vartheta(t_{k-1})$ are the
effective rotation angles applies during the $k$th square pulse.  At
this point, we may interpret each of the terms of
\eqn{eq:corpse-terms} as vectors on the dynamical Lie algebra.  Figure
\ref{fig:corpse}a is a diagram of these vectors on $\su{2}$.  In order
to eliminate the first-order expansion term, the rotation angles
$\theta_k$ must be chosen so that the vectors must form a closed path.

Then by choosing the rotation angles to be
\begin{eqnarray}
    \theta_1 &=& 2\pi n_1 + \theta/2 - \arcsin \left( \sin(\theta/2)/2 \right) \nonumber \\ 
    \theta_2 &=& 2\pi n_2 - 2 \arcsin \left( \sin(\theta/2)/2 \right) \nonumber \\
    \theta_3 &=& 2\pi n_3 + \theta/2 - \arcsin \left( \sin(\theta/2)/2 \right),
\label{eq:corpse-thetas}
\end{eqnarray}
where $n_1$, $n_2$ and $n_3$ are integers, we find that both $\ep{D}
\Omega_1(\tau,0) = 0$ and $\theta = (\theta_1 - \theta_2 + \theta_3
\mod 2\pi)$.  The extra factors of $2\pi$ are added so that the
individual pulse rotation angles may be made positive.  In principle,
any choice of integers is sufficient to compose a first-order
sequence, however the choice $n_1 = n_2 = 1$ and $n_3=0$ minimizes the
remaining second-order term while still producing a positive set of
rotation angles \cite{Cummins2003}.  The CORPSE family of sequences,
\begin{eqnarray}
V(\u(t)) &=& \exp(-\i (\theta_3 \Hx + \ep{D} \theta_3 \Hz)) \exp(-\i(-\theta_2 \Hx + \ep{D} \theta_2 \Hz)) \exp(-\i(\theta_1 \Hx + \ep{D} \theta_1 \Hz)) \nonumber \\
M_{\mathrm{C1}}(\theta,0) &=& M(\theta_3, 0) M(\theta_2, \pi) M(\theta_1, 0)  = R(\theta,0) + \O(\ep{D}^2)
\label{eq:corpse}
\end{eqnarray}
are fully compensating first-order sequences.  In the presence of an
unknown detuning error, a CORPSE sequence,
$M_{\mathrm{C1}}(\theta,0)$, may be implemented in the place of the
simple rotation $R(\theta,0)$ and the erroneous evolution is
suppressed.  In figure \ref{fig:corpse}b, we plot the magnetization
trajectory for a qubit under a CORPSE sequence with detuning error
$\ep{D} = 0.2$.

\subsubsection{Arbitrarily accurate CORPSE}
We turn our attention to sequences which compensate detuning errors to
arbitrarily high order.  Once again, the Solovay-Kitaev method may be
used to construct higher-order sequences, now using CORPSE as the seed
sequence.  This problem was first studied by Alway and Jones
\cite{Alway2007}.  Recall that in the Solovay-Kitaev method, one
synthesizes a correction sequence $B_{n+1}$ in two steps.  First, one
generates a propagator $P_{(n+1)z}(\xi)$ with an amplitude
proportional to the leading-order error term.  Second, the similarity
transformation $U_T \Upsilon^\dagger$ is used so that $B_{n+1} = \U_T
\Upsilon^\dagger P_{(n+1)z}(\xi) \Upsilon U^\dagger_T$ cancels the
leading error term of the seed sequence (see \eqn{eq:Bn}).
\eqn{eq:recursive-Pj} shows that the term $P_{(n+1)z}(\xi)$ may be
constructed recursively using a product of first-order $P_1$'s.
Therefore the Solvay-Kitaev method may be extended to correct detuning
errors if: (1) procedures for generating $P_{1x}$, $P_{1y}$, and
$P_{1z}$ using imperfect pulses have been found, and (2) when a method
for applying the similarity transformation $U_T \Upsilon^\dagger$
using imperfect pulses has been identified.  This task is complicated
by the difficulty of generating inverse operations for imperfect
pulses affected by detuning errors.

Earlier, the first-order $P_{1}$'s were applied using a composite
pulse sequence (see \eqn{eq:Sx-sequence}).  We employ a similar
strategy here.  Let 
\begin{eqnarray}
S_z(\alpha) = M(\alpha/2,\phi+\pi) M(\alpha/2,\phi) = \exp(-\i \alpha \epsilon \Hz) = P_{1z}(\alpha).
\end{eqnarray}
Then we may apply $P_{1z}(\alpha)$ by implementing the sequence
$S_z(\alpha)$ instead.  What remains is to develop sequences that
apply $P_{1x}(\alpha)$ and $P_{1y}(\alpha)$.  Observe that if it were
possible to perform ideal rotations (where $R^\dagger(\theta,\phi) =
R(\theta,\phi + \pi)$), then one could easily implement $S_x(\alpha) =
R(\pi/2, \pi/2) S_z(\alpha) R^\dagger(\pi/2, \pi/2)$ and $S_y(\alpha)
= R^\dagger(\pi/2, 0) S_z(\alpha) R(\pi/2, 0)$ in the place of
$P_{1x}(\alpha)$ and $P_{1y}(\alpha)$.  However in the presence of
systematic detuning errors, we may not apply this transformation using
a simple imperfect pulse since $M^\dagger(\theta,\phi) \neq
M(\theta,\phi + \pi)$.  We may avoid this complication by using a
CORPSE sequence to approximate the transformation to sufficient
accuracy.  We use the first-order CORPSE sequence, $M_{\mathrm{C1}}(\theta,\phi)$
to implement the target operation $U_T = R(\theta,\phi)$.  Then we may
write
\begin{eqnarray}
S_x(\alpha) = M_{\mathrm{C1}}(\pi/2,\pi/2) S_z(\alpha) M_{\mathrm{C1}}(\pi/2, 3\pi/2) = \exp(-\i \alpha \epsilon \Hx) + \O(\epsilon^2) = P_{1x}(\alpha)
\end{eqnarray}
and
\begin{eqnarray}
S_y(\alpha) = M_{\mathrm{C1}}(\pi/2,\pi) S_x(\alpha) M_{\mathrm{C1}}(\pi/2,0) = \exp(-\i \alpha \epsilon \Hy) + \O(\epsilon^2) = P_{1y}(\alpha).
\end{eqnarray}
Furthermore, by careful choice of $\theta$ and $\phi$, we can
implement $P_{1\eta}(\alpha)$ for any axis $\eta$ about any angle
$\alpha$. Following the Solovay-Kitaev method (Sec. \ref{SK-method}),
we can then construct $P_{jz}(\xi)$ to any order $j$.

What remains is to implement the similarity transformation by $U_T
\Upsilon^\dagger$. However we note that the available approximate
rotations $M_{\mathrm{C1}}(\theta,\phi) = R(\theta,\phi) + \O(\ep{D})$
are only accurate to first order.  Hence, if we were to attempt to
apply a similarity transformation on a higher-order $P_{jz}(\xi)$ term
using CORPSE sequences, additional second-order errors would be
introduced.  This difficulty may be avoided by applying the
appropriate rotation at the level of the first-order $P_1$
operations. As an instructive example, we work through the process for
the second-order correction.  The correction sequences may be
constructed in a rotated coordinate system, determined by the basis
transformation $H_\mu^\prime = U_T\Upsilon^\dagger H_\mu\Upsilon
U_T^\dagger$ for $\mu \in \{ \mathrm{x}, \mathrm{y}, \mathrm{z} \}$.
One may calculate four rotations $R(\theta_{\pm\mu^\prime},
\phi_{\pm\mu^\prime})$, that map the $\Hz$ axis to the $H_{\pm
  \mathrm{x}^\prime}$ and $H_{\pm \mathrm{y}^\prime}$ axes
$H_{\pm\mu^\prime}=R(\theta_{\pm\mu^\prime},\phi_{\pm\mu^\prime})\Hz
R^\dagger(\theta_{\pm\mu^\prime},\phi_{\pm\mu^\prime})$ and also $P_{1
  \pm \mu^\prime}(\alpha) =
R(\theta_{\pm\mu^\prime},\phi_{\pm\mu^\prime}) P_{1z}(\alpha)
R^\dagger(\theta_{\pm\mu^\prime},\phi_{\pm\mu^\prime})$.  This
transformation may also be applied using a first-order CORPSE
sequence, since the resulting error is absorbed into the second-order
error term in $P_{1 \pm \mu^\prime }(\alpha)$.  Once transformed terms
have been obtained, the correction sequence may be constructed in the
usual manner
\begin{eqnarray*}
B_2 &=& P_{2z^\prime}(\xi) = P_{1x^\prime}(-\sqrt{\xi})P_{1y^\prime}(-\sqrt{\xi})P_{1x^\prime}(\sqrt{\xi})P_{1y^\prime}(\sqrt{\xi}) \\
&=& M_{\mathrm{C1}}(\theta_{-x^\prime},\phi_{-x^\prime})P_{1z}(\sqrt{\xi})M_{\mathrm{C1}} (\theta_{-x^\prime},\phi_{-x^\prime}+\pi)M_{\mathrm{C1}}(\theta_{-y^\prime},\phi_{-y^\prime})P_{1z}(\sqrt{\xi})M_{\mathrm{C1}} (\theta_{-y^\prime},\phi_{-y^\prime}+\pi)\\
&\:& \times M_{\mathrm{C1}}(\theta_{x^\prime},\phi_{x^\prime})P_{1z}(\sqrt{\xi})M_{\mathrm{C1}}(\theta_{x^\prime},\phi_{x^\prime}+\pi)M_{\mathrm{C1}}(\theta_{y^\prime},\phi_{y^\prime}) P_{1z}(\sqrt{\xi})M_{\mathrm{C1}}(\theta_{y^\prime},\phi_{y^\prime}+\pi).
\end{eqnarray*}
We can then define a second order CORPSE sequence as
$M_{\mathrm{C2}}(\theta,0)$$=B_2M_{\mathrm{C1}}(\theta,0)$$=R(\theta,0)+\O(\ep{D}^3)$. One
can then continue using Solvay-Kitaev techniques to remove the
detuning error to all orders.  


\subsubsection{Concatenated CORPSE: correcting simultaneous errors}
\label{simultaneous1}
So far, we have considered the problem of quantum control in the
presence of a single systematic error, while in a real experiment
several independent systematic errors may affect the controls.  In
many situations one error dominates the imperfect evolution; it is
appropriate in these cases to use a compensating sequence to suppress
the dominant error.  However it is also important to consider whether
a sequence reduces the sensitivity to one type of error at the cost of
increased sensitivity to other types of errors \cite{Cummins2003}.
Such a situation may occur if an error model couples two error
sources.  Consider a set of control functions $\{u_\mu(t)\}$ where
each control is deformed by two independent systematic errors.  A
general two-parameter error model for the control $v_\mu(t) =
f_\mu[\u(t);\epsilon_i , \epsilon_j]$ may be formally expanded as
\begin{eqnarray}
f_\mu[ \u(t); \epsilon_i , \epsilon_j ] &=& f_\mu[\u(t);0,0] + \epsilon_i \frac{\partial}{\partial \epsilon_i}  f_\mu[\u(t);0,0] + \epsilon_j \frac{\partial}{\partial \epsilon_j} f_\mu[\u(t);0,0] + \nonumber \\ 
&\,& 2 \epsilon_i \epsilon_j \frac{\partial^2}{\partial \epsilon_i \partial \epsilon_j } f_\mu[\u(t);0,0] + \O(\epsilon_i^2 + \epsilon_j^2 ).
\label{eq:two-error-expansion}
\end{eqnarray}
In practice, we need only concern ourselves with the first few
expansion terms, since in most physically relevant error models the
higher order derivatives are identically zero.  Following the same
reasoning employed in section \ref{errors}, we may decompose the
imperfect propagator as $V(\u(t)) = U(\u(t))U^I(\vec{v}(t) - \u(t))$,
where $U(\u(t))$ is the ideal propagation in the absence of errors and
the interaction frame propagator $U^I(\vec{v}(t) - \u(t))$ represents
the evolution induced by the systematic errors.  Formally
$U^I(\vec{v}(t) - \u(t))$ may be studied using a Magnus expansion,
however this method is usually impeded by the complexity of the Magnus
series.  The determination of sequences which compensate simultaneous
errors is currently an unresolved problem, although some progress has
been made by considering concatenated pulse sequences
\cite{Ichikawa2011b}.  As an example relevant to an NMR quantum
computer, we now study error models where two systematic errors occur
simultaneously and show that these errors can be compensated by
concatenation of pulse sequences.

\emph{Simultaneous amplitude and detuning errors:} When studying the
control of qubits based on coherent spectroscopy methods, it is
natural to consider situations where systematic errors in the field
amplitude and tuning are simultaneously present.  From the NMR control
Hamiltonian \eqn{eq:nmr} it is straightforward to derive the error
model, $v_\mathrm{x/y} = u_\mathrm{x/y}(t) ( 1 + \ep{A} )$ and
$v_\mathrm{z} = \uz(t) + \ep{D}$.  In this case, the independent
amplitude and detuning errors are decoupled and the parameters
$\ep{A}$ and $\ep{D}$ do not affect the same control.
    
\emph{Simultaneous pulse-length and detuning errors:} Likewise,
simultaneous errors in the pulse length and field tuning may occur.
Again, it is simple to derive the joint error model, $v_\mathrm{x/y} =
u_\mathrm{x/y}(t) ( 1 + \ep{T} )$ and $v_\mathrm{z} = (\uz(t) +
\ep{D}) (1 + \ep{T})$.  Unlike the previous case, this error model
couples the parameters $\ep{T}$ and $\ep{D}$, since they both effect
the $\Hz$ control function.

\begin{figure}
\begin{center}
\includegraphics{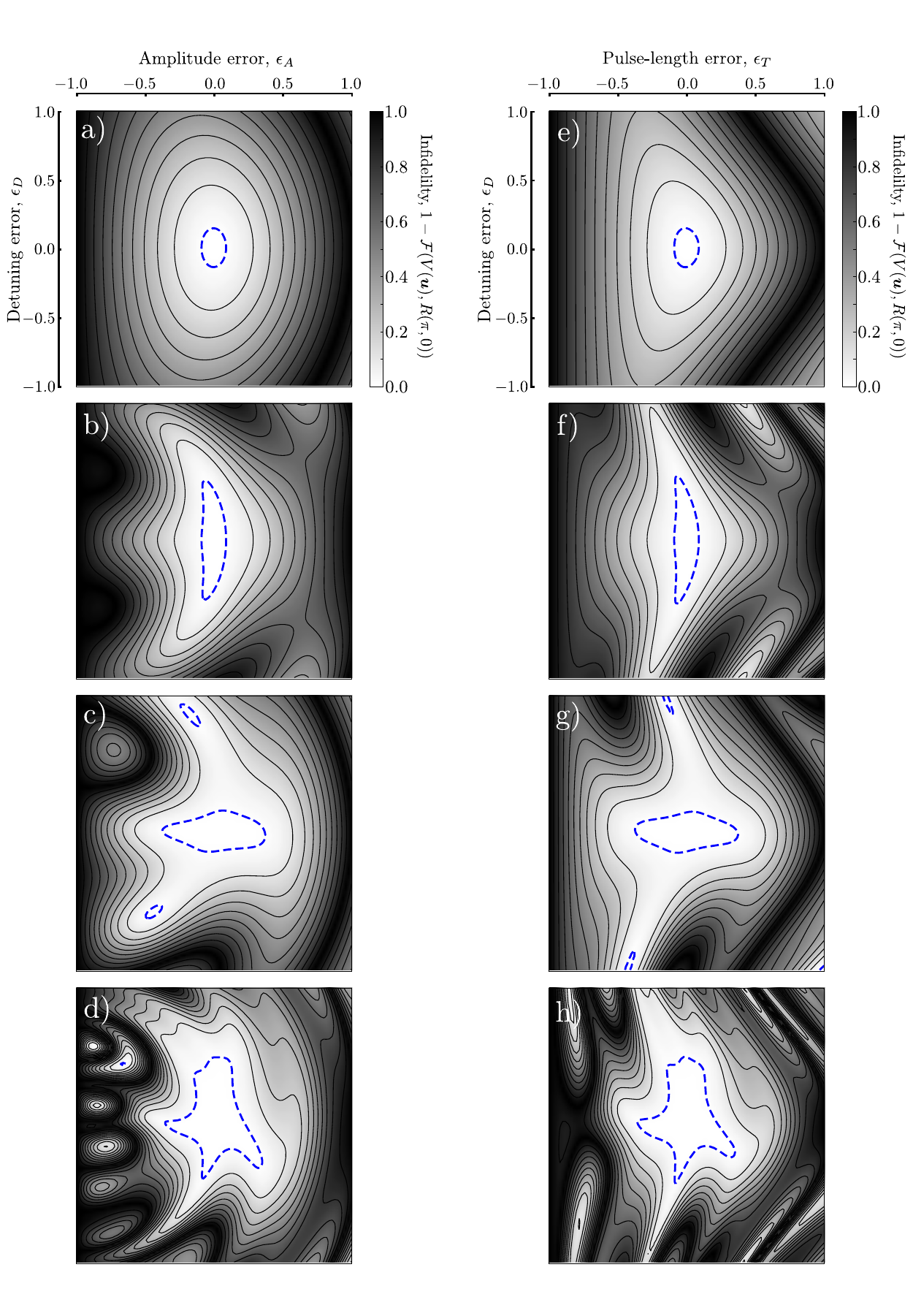}
\end{center}
\caption{Infidelity of a,e) Plain pulses, b,f) CORPSE  c,g)
  B2, and d,h) a concatenated B2CORPSE sequence in the
  presence of simultaneous amplitude and detuning (left column) and
  pulse-length and detuning (right column) errors.  The target
  rotation is $U_T = R(\pi,0)$.  The dashed contour corresponds to an
  infidelity of 0.01, while the remaining contours are plotted at 10\%
  intervals. \label{B2CORPSE}}
\end{figure}

It is sometimes possible to produce a sequence that compensates two
simultaneous errors by pulse sequence concatenation.  Recall from
section \ref{wimperis} that the Wimperis sequence B2 is a second-order
compensation sequence for both amplitude and pulse-length errors.
However if only detuning errors are present, then the response of a B2
sequence to pure detuning errors is similar to that of a native pulse
Similarly, a CORPSE sequence would eliminate
the first order term produced by the detuning offsets but not correct amplitude errors.
The sequence B2CORPSE corrects both
errors \emph{independently} by concatenating the B2 and CORPSE
sequences.  B2CORPSE is composed of three B2 subsequences that form a
larger CORPSE sequence
\begin{eqnarray}
M_{\mathrm{B2C1}}(\theta,0) = M_{\mathrm{B2}}(\theta_3,0)M_{ \mathrm{B2}}(\theta_2,\pi)M_{ \mathrm{B2}}(\theta_1,0),
\end{eqnarray}
where the angles $\{\theta_1, \theta_2, \theta_3\}$ are given in
\eqn{eq:corpse-thetas}.  At the lower level of concatenation, B2
sequences are used to synthesize rotations robust to pure amplitude or
pulse-length errors, whereas on the higher level the CORPSE
construction compensates pure detuning errors.  Figures
\ref{B2CORPSE}a-d are plots of the infidelity of plain pulses, and the
CORPSE, B2, and B2CORPSE sequences in the presence of simultaneous
amplitude and detuning errors.  Similarly, figures \ref{B2CORPSE}e-h
are infidelity plots for simultaneous pulse-length and detuning
errors.  As expected, the CORPSE sequences improve the accuracy of
gates with respect to detuning errors, but offers little improvement
against either amplitude or pulse length errors.  Unsurprisingly, this
behavior is inverted for the B2 sequences.  The B2CORPSE sequence
performs well for either amplitude/pulse-length or detuning errors; in
the presence of simultaneous errors the performance diminishes, yet
the fidelity of the applied gate is still vastly improved over
uncompensated pulses.


\subsection{Shaped pulse sequences}
\label{shaped}
Thus far, we have considered sequences composed control functions
$u_\mu(t)$ which are piecewise constant over the pulse interval.  This
construction is particularly convenient from a sequence design
perspective, although in practice instrumental shortcomings will
frequently distort the pulse profile.  Also, in some applications, the
rectangular profile is non-ideal.  For example, in the frequency domain
the square pulse corresponds to a sinc function, whose local maxima
may complicate the control of certain systems.  Finally, we note that
in some systems, such as in superconducting Josephson junction qubits,
bandwidth requirements forbid the sudden switching of control fields
(i.e., place a limit on the derivative $|\dot{u}_\mu| \leq
\dot{u}_\mu^{\mathrm{MAX}}$).  In these cases, it is desirable to
consider shaped pulse sequences composed by a set of continuous
differentiable control functions \cite{Steffen2003}.  
In the present article, we shall only consider shaped pulse sequences
which are also fully-compensating (class A), and therefore appropriate
for use in a quantum processor.  Specifically, we study shaped pulse
sequences which also compensate systematic errors.  We shall see that
the additional flexibility admitted by shaped pulses frequently
produces superior sequences.

Once given a pulse waveform, we may verify that it is compensating for
a particular error model by computing the Magnus expansion in the
appropriate interaction frame (see section \ref{method}); however, we
emphasize these methods require the evaluation of successive nested
integrals (e.g., \eqn{eq:magnus}), and are difficult to analytically
implement beyond the first few orders \cite{Pryadko2008}.
Furthermore, the inverse problem (solving for the control functions)
is especially difficult except for in the most simple cases.  For this
reason, various numerical optimization methods have become popular,
including gradient-accent techniques, optimal control methods
\cite{Skinner2003, Skinner2004, Li2011}, and simulated annealing
\cite{Geen1989,Geen1991}.  Methods which use elements of
optimal-control theory merit special attention; in recent years the
GRAPE \cite{Khaneja2005, DeFouquieres2011} and Krotov algorithms have
been especially successful in pulse design, and has been applied to
NMR \cite{Skinner2003}, trapped ions \cite{Timoney2008}, and ESR
\cite{Hodges2008}.  The main advantage of the GRAPE algorithm is an
efficient estimation of the gradient of the fidelity as function of
the controls, which then enables optimization via a gradient-accent
method.  A recent review of these methods may be found in Refs.
\cite{Singer2010,Machnes2011}.

To demonstrate the relative performance of shaped sequences, we
consider the continuous analogs of the CORPSE sequence that modulate
$\ux(t)$ to compensate detuning errors.  
From \eqn{eq:CORPSE-magnus} observe that the first-order compensation
condition $\ep{D} \Omega_1(\tau,0) = 0$ is met when $\int_0^{\tau} dt
\cos( \vartheta(t) ) = \int_0^{\tau} dt \sin( \vartheta(t) ) = 0$.
What remains is to solve for a set of control functions $\ux(t)$ which
generate the target operation $U_T$ while remaining robust to the
detuning error. One popular scheme is to decompose the control
functions as a Fourier series \cite{Geen1991,Pryor2007} over the pulse
interval, 
\begin{eqnarray}
  \ux(t) = \omega \sum_{n=0}^{\infty} a_n \cos \left(n \omega \left(t - \frac{\tau}{2}\right) \right) + b_n \sin \left(n \omega \left(t - \frac{\tau}{2} \right) \right), \qquad \omega = \frac{2 \pi}{\tau}.
\end{eqnarray}
This has the advantage of specifying the controls using only a few
expansion parameters $\{a_n\}$ and $\{b_n\}$.  Also, the series may be
truncated to avoid high frequency control modulations that are
incompatible with some control systems.  Moreover, by choosing each of
the expansion terms $b_n = 0$, the resultant sequence may be made
symmetric with respect to time reversal.  We note that not all
compensating sequences completely eliminate the first order error
term; in some sequences the leading order errors are highly suppressed
rather than completely eliminated.  This behavior is more common in
sequences obtained from numerical methods.

\begin{figure}
\begin{center}
\includegraphics{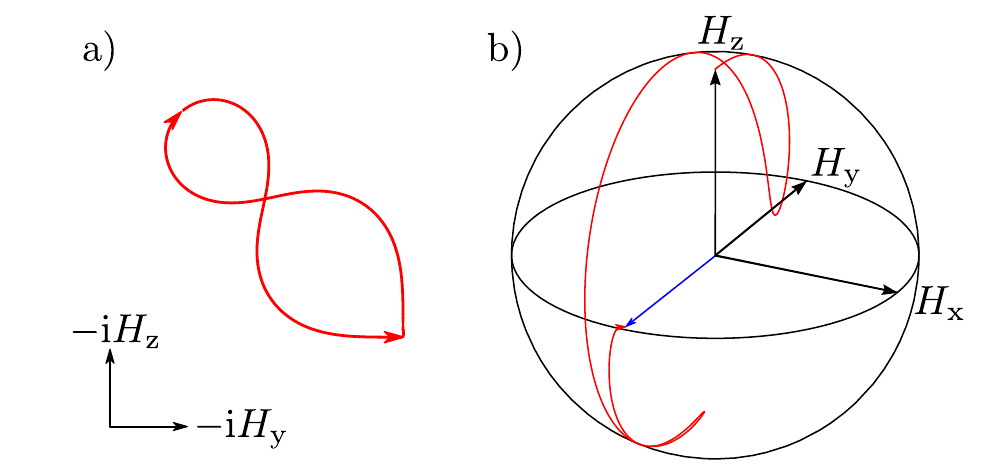}
\end{center}
\caption{a) Vector path on the Lie algebra traced out by the
    interaction frame Hamiltonian $H^I(t) = \ep{D} \Hz^I(t)$ for the
    shaped sequence S$_1(\pi/2)$.  b) Magnetization trajectory for a
    qubit under an S$_1(\pi/2)$ sequence for the target operation $U_T
    = R(\pi/2,0)$, for the exceptionally large detuning error
    $\ep{D} = 1$. \label{fig:S1}}
\end{figure}

\begin{figure}
\begin{center}
\includegraphics{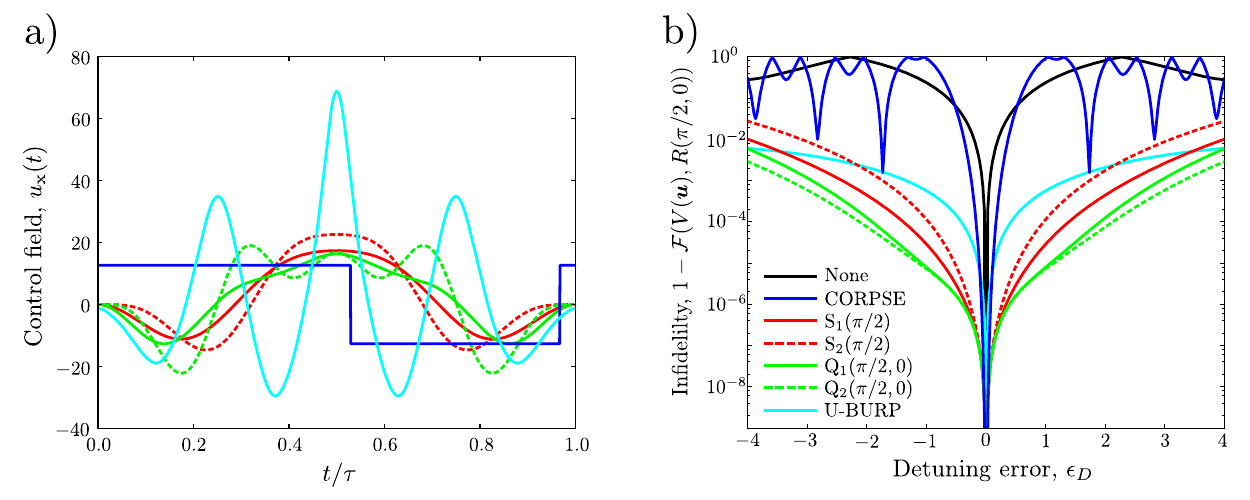}
\end{center}
\caption{a) Control function $\ux(t)$ for various pulse sequences
  designed to compensate systematic detuning errors.
  b) Performance of several shaped pulses over a wide detuning range. \label{fig:shaped-scaling}}
\end{figure}

Historically, among the first class A shaped pulses developed belonged
to the U-BURP (band-selective, uniform response, pure-phase) family
designed by Geen and Freeman \cite{Geen1991, Freeman1998}, using
simulated-annealing and gradient decent methods.  Soon after,
Abramovich and Vega attacked the same problem using approximation
methods based on Floquet theory \cite{Abramovich1993}.  These
sequences are designed to correct detuning errors over an extremely
large range; however, they fail to completely eliminate the leading
order error terms.  In the design of compensating sequences for
quantum computing, the emphasis has been on the synthesis of composite
rotations of extraordinary accuracy over a narrow window of errors.
More recently, Steffen and Koch considered shaped Gaussian modulated
controls specially designed for superconducting qubit manipulations
\cite{Steffen2007}.  Pryadko and Sengupta also derived a family of
sequences (S$_1(\phi_0)$,S$_2(\phi_0)$,Q$_1(\phi_0)$,Q$_2(\phi_0)$)
using a semi-analytic method based on the Magnus expansion and average
Hamiltonian theory \cite{Sengupta2005,Pryadko2008}.  By design, these
sequences eliminate the leading-order error terms in a manner similar
to the square-pulse sequences previously discussed.

It is interesting to compare the performance of shaped sequences to a
sequence of square pulses, such as CORPSE.  Many of these sequences
compensate detuning errors by modulating the amplitude of the control
function $\ux(t)$.  In figure \ref{fig:shaped-scaling}a pulse shapes
are given for CORPSE, U-BURP, S$_1(\pi/2)$, S$_2(\pi/2)$,
Q$_1(\pi/2)$, and Q$_2(\pi/2)$ sequences that implement the target
operation $U_T = R(\pi/2,0)$.  The relative performance of these
sequence over a wide range of systematic detuning error is given in
figure \ref{fig:shaped-scaling}b.  Clearly, shaped pulses outperform
CORPSE over a wide range of field tunings; however, for very small
detunings, CORPSE has more favorable scaling behavior.  Figures
\ref{fig:S1}a and \ref{fig:S1}b show the behavior of the shaped pulse
S$_1(\pi/2)$ as trajectories on the Lie algebra and on the Bloch
sphere respectively.


\section{Composite pulse sequences on other groups}
We have studied sequences which compensate imperfect single-qubit
rotations, i.e., operations which form a representation of the group
$\SU{2}$.  Another class of problems of practical and fundamental
interest is the design of sequences for other Lie groups, such as the
group of $n$-qubit operations $\SU{2^n}$ \cite{Khaneja2001}.  Several
compensating sequences exist for multi-qubit gates
\cite{Jones2003,Testolin2007,Tomita2010}.  In general, these sequences
rely on operations which form an $\SU{2}$ subgroup to perform
compensation in a way analogous to the one-qubit case and is the 
topic of this section.  

We note the design of compensating pulse sequences that do not rely on an
$\SU{2}$ or $\group{SO}(3)$ subgroup is a largely unexplored topic,
and is an interesting subject for future study.  For many cases, we 
can determine which errors can be compensated by examining the algebra \cite{Li2006}, 
but it is unclear what the natural compensating sequences are for groups 
that are not equivalent to rotations in three-dimensions.


\subsection{Compensated two-qubit operations}
\label{two-qubit}
Thus far, we have shown how compensating sequences may be used to
correct systematic errors in arbitrary one-qubit operations.
Universal quantum computation also requires accurate two-qubit gates.
The study of two-qubit compensating sequences is therefore of great
fundamental and practical interest.  In this section, we study
two-qubit operations using Lie theoretic methods.  The Cartan
decomposition of the dynamical Lie algebra will be central to this
approach.  We begin by studying the properties and decompositions of
the Lie group and its associated algebra.  Then we consider systematic
errors in two-qubit gates derived from the Ising interaction, and show
how the Cartan decomposition may be used to construct compensating
sequences.

Any two-qubit Hamiltonian (up to a global phase) may be written
in the form,
\begin{eqnarray}
H(t) = \sum_{\mu} \sum_{\nu} u_{\mu \nu} (t) H_{\mu \nu},
\label{eq:two-qubit-hamiltonian}
\end{eqnarray}
where the controls are represented in the product-operator basis
\cite{Sorensen1983}, corresponding to the control Hamiltonians $H_{\mu
    \nu} = 2 H_\mu \otimes H_\nu$, where $\H{1} = \frac{1}{2} \Id$ and
the indices $\mu$, $\nu$ run over $H_\mu \in \{ \H{1}, \Hx, \Hy, \Hz
\}$.  As a matter of convention, we exclude the term proportional to
the identity $H_{11} = \frac{1}{2} \Id \otimes \Id$ since it generates
an unimportant global phase and otherwise does not contribute to the
dynamics; it is implied in the following equations that this term
never appears.  The product-operator representation is particularly
convenient since each control Hamiltonian is orthogonal under the
Hilbert-Schmidt inner product, $\iprod{H_{\mu \nu}}{H_{\rho
        \sigma}} = \delta_{\mu,\rho} \delta_{\nu,\sigma}$.  The
family of all possible solutions to a control equation, for example
\eqn{eq:control} with the Hamiltonian \eqn{eq:two-qubit-hamiltonian},
forms a representation of the special unitary group $\SU{4}$.  This
group contains all possible single-qubit (local) operations that may
be applied to each qubit among the pair as well as all two-qubit
(nonlocal) operations.

\emph{Cartan decomposition of two-qubit gates:} Associated with the
group $\SU{4}$ is the corresponding Lie algebra $\su{4} =
\bigoplus_{\mu \nu} \span\{-\i H_{\mu \nu}\}$ excluding $\H{11}$
(including $\H{11}$ would make the group $\group{U}(4)$).
Consequently $\su{4}$ is the algebra of four-dimensional traceless
skew-Hermitian matrices.  Observe that $\su{4}$ may be decomposed as
$\su{4} = \algb{k} \oplus \algb{m}$, where
\begin{eqnarray*}
\algb{k} &=& \su{2} \otimes \su{2} = \span \{-\i \H{x1},-\i \H{y1},-\i \H{z1},-\i \H{1x},-\i \H{1y},-\i \H{1z}\} \\
\algb{m} &=& \span \{-\i \H{xx}, -\i \H{xy}, -\i \H{xz}, -\i \H{yx}, -\i \H{yy}, -\i \H{yz}, -\i \H{zx}, -\i \H{zy}, -\i \H{zz} \}.
\end{eqnarray*}
It may be verified that $[\algb{k},\algb{k}] \subseteq \algb{k}$,
$[\algb{m},\algb{k}] = \algb{m}$, and $[\algb{m},\algb{m}] \subseteq
\algb{k}$.  Therefore, the decomposition $\algb{k} \oplus \algb{m}$ is
a Cartan decomposition of the algebra $\su{4}$ (see section
\ref{cartan}).  Also, the subalgebra $\algb{a} = \span\{ -\i \H{xx},
-\i \H{yy}, -\i \H{zz} \}$ is a maximal abelian subalgebra and
$\algb{a} \subset \algb{m}$.  Therefore, $\algb{a}$ is a Cartan
subalgebra.  This admits the decomposition $U = K_2 A K_1$ for any
element $U \in \SU{4}$.  The operators
\begin{eqnarray}
K_j = \exp \left( -\i  \sum_{\mu \in \{\mathrm{x,y,z}\}} \alpha^{(j)}_{\mathrm{\mu 1}} \H{\mu1} + \alpha^{(j)}_{\mathrm{1 \mu}} \H{1\mu} \right) 
\end{eqnarray}
are in the subgroup $\e^{\algb{k}} = \SU{2} \otimes \SU{2}$ comprising
all single-qubit operations for both qubits, whereas the
operator
\begin{eqnarray}
A = \exp \left( -\i \sum_{ \mu \in \{\mathrm{x,y,z} \}} \alpha_{\mu \mu} \H{\mu \mu} \right)
\end{eqnarray}
is in the abelian group $\e^{\algb{a}} = \group{T}^3$, isomorphic to
the 3-torus, generated by the two-qubit interaction terms
$\H{xx}$, $\H{yy}$, and $\H{zz}$.  In analogy with the Euler
decomposition, the parameters $\alpha_{\mu \nu}$ may be regarded as
rotation angles.  This construction is a $KAK$ Cartan decomposition
for the propagator $U$, and may be used as a framework for a pulse
sequence to generate any arbitrary two-qubit operation.  Furthermore,
if the operators $K_2$, $A$, and $K_1$ may be implemented using a
compensating pulse sequence, robust for a given error model, then
their product will also be a compensating sequence.


\subsubsection{Operations based on the Ising interaction:} We seek a
method for implementing an arbitrary gate $A \in \e^{\algb{a}}$ using
a two-qubit coupling interaction.  To be specific, we consider an NMR
quantum computer of two heteronuclear spin qubits coupled by an Ising
interaction.  It is well known that propagators generated by the Ising
interaction and single-qubit rotations are universal for $\SU{4}$
\cite{Nielsen2000}; for completeness we explicitly show how
arbitrary gates may be synthesized using the $KAK$ form.  Under the
assumption of the qubits can be spectrally distinguished due to large differences 
in Larmor frequencies, the system Hamiltonian takes the form
\begin{eqnarray}
H(t) = \left( \sum_{\mu \in \{\mathrm{x,y,z}\} } u_{\mu1}(t) H_{\mu1} + u_{1\mu}(t) H_{1\mu} \right ) + \uu{zz}(t) \H{zz},
\label{eq:ising-hamiltonian}
\end{eqnarray}
where the controls $u_{\mu1}(t)$ and $u_{1\mu}(t)$ are applied by the
appropriate rf fields (see section \ref{nmr}) and $\uu{zz}(t) = 2
\pi J(t)$ is the strength of the spin-spin coupling interaction.  In
some cases, it is useful to manipulate the strength of the scalar
coupling $J(t)$, e.g., by using spin decoupling techniques
\cite{Freeman1998}.  We consider the simpler case of constant Ising
couplings, and also assume that it is possible to apply hard pulses,
where rf coupling amplitude greatly exceeds $J$, and the spin-spin
coupling may be considered negligible.  In terms of resource
requirements for the quantum computer, this implies that single qubit
operations are fairly quick, whereas two-qubit operations driven by
the Ising coupling are much slower.

Consider the propagators $U_{\mu \mu}(\alpha) = \exp( - \i \alpha
\H{\mu \mu} ) \in \e^{\algb{a}}$.  Let us define the one-qubit
rotation operators $R_1(\theta,\phi) = R(\theta,\phi) \otimes \Id$ and
$R_2(\theta,\phi) = \Id \otimes R(\theta,\phi)$. In the absence of
applied rf fields, the system evolves according to
$U_{\mathrm{zz}}(\theta_{\mathrm{zz}})$, where $\theta_{\mathrm{zz}} =
\uu{zz} \Delta t$ and $\Delta t$ is the duration of the
free-precession interval.  Observe that for
\begin{eqnarray}
K_\mathrm{x} &=& R_1(\pi/2, 0) R_2(\pi/2, 0) = \exp(-\i \pi/2 ( \H{x1} + \H{1x} ) ) \nonumber \\
K_\mathrm{y} &=& R_1(\pi/2, \pi/2) R_2(\pi/2, \pi/2) = \exp(-\i \pi/2 ( \H{y1} + \H{1y} ) ),
\end{eqnarray}
$U_{\mathrm{xx}}(\alpha) = K_\mathrm{y}
U_\mathrm{zz}(\alpha) K^\dagger_\mathrm{y}$ and
$U_{\mathrm{yy}}(\alpha) = K^\dagger_\mathrm{x} U_\mathrm{zz}(\alpha)
K_\mathrm{x}$.  Since the group $\e^{\algb{a}}$ is abelian, then any
arbitrary group element $A$ specified by the decomposition angles
$\{\alpha_{\mathrm{xx}} , \alpha_{\mathrm{yy}},\alpha_{\mathrm{zz}}
\}$, may be produced by the product $A =
U_{\mathrm{xx}}(\alpha_{\mathrm{xx}})
U_{\mathrm{yy}}(\alpha_{\mathrm{yy}})U_{\mathrm{zz}}(\alpha_{\mathrm{zz}})$.
Then any $U = K_2 A K_1 \in \SU{4}$ may be produced using single qubit
rotations and the Ising interaction.

\emph{Systematic errors in Ising control:} In practice, systematic
errors introduced by experimental imperfections prohibit the
application of perfect two-qubit gates.  In the case of gates produced
by the Ising interaction, the errors may arise from several sources,
such as experimental uncertainty in the strength of the coupling $J$.
It is therefore desirable to design sequences that implement accurate
Ising gates over a range of coupling strengths.  The error model in
this case is similar to the case of amplitude errors; the control
$\uu{zz}$ is replaced by the imperfect analogue $v_\mathrm{zz} =
\uu{zz}(1 + \ep{J})$ where the parameter $\ep{J}$ is proportional
to the difference between the nominal (measured) and actual coupling
strengths.  For now we assume the remaining controls are error free.
In correspondence, the perfect propagators
$U_\mathrm{zz}(\theta_\mathrm{zz})$ are replaced by
$V_\mathrm{zz}(\theta_\mathrm{zz}) =
U_\mathrm{zz}(\theta_\mathrm{zz}(1 + \ep{J}))$.

Jones was the first to study compensating pulse sequences for the
Ising interaction \cite{Jones2003, Xiao2006}, and proposed sequences
closely related to the Wimperis sequences studied in section
\ref{wimperis}.  Here we demonstrate that these sequences are easily
derived using Lie algebraic techniques.  Observe that there are
several subalgebras contained in $\su{4}$, for instance $\algb{j} =
\span \{ -\i \H{x1}, -\i \H{yz}, -\i \H{zz} \}$ that are representations of $\su{2}$.  
Furthermore, if it is
feasible to produce any imperfect propagator in the group $\group{J} =
\e^\algb{j}$, then the compensating sequences discussed in section
\ref{SU2} may be reused for this system.

This strategy can be implemented using
accurate one-qubit rotations.  For example, let
\begin{eqnarray}
    \mathcal{R}(\theta,\phi) =
     R_1^\dagger(\phi,0) U_\mathrm{zz}(\theta) R_1(\phi,0)
     = \exp(-\i \theta ( \cos \phi \H{zz} + \sin \phi \H{yz})).
\label{eq:ising-rotation}
\end{eqnarray}  
Also, let us define the imperfect rotation $\mathcal{M}(\theta,\phi) =
R_1^\dagger(\phi,0) V_\mathrm{zz}(\theta) R_1(\phi,0) =
\mathcal{R}(\theta(1+\ep{J}),\phi)$.  The two qubit unitaries
$\mathcal{R}(\theta,\phi)$ are isomorphic to single qubit rotations
$\mathcal{R}(\theta,\phi)$ and Jones used this similarity to construct
an alternative sequence we refer to as B2-J \cite{Jones2003,
    Tomita2010},
\begin{eqnarray}
   \mathcal{M}_{\text{B2-J}}(\theta,0) &=& \mathcal{M}(\pi,\phi) \mathcal{M}(2\pi, 3\phi) \mathcal{M}(\pi,\phi) \mathcal{M}(\theta,0) \nonumber \\
&=& R_1^\dagger(\phi, 0)U_\mathrm{zz}(\pi(1+\ep{J})) R_1(\phi,0)  \: R_1^\dagger(3\phi, 0)U_\mathrm{zz}(2\pi(1+\ep{J})) R_1(3\phi,0) \: \times \nonumber \\ &\:& R_1^\dagger(\phi, 0)U_\mathrm{zz}(\pi(1+\ep{J})) R_1(\phi,0)\:U_\mathrm{zz}(\theta(1+\ep{J})) \nonumber \\ &=& U_\mathrm{zz}(\theta) + \O(\ep{J}^3)
\end{eqnarray}
where again $\phi = \arccos( - \theta / 4 \pi )$.  If the sequence
B2-J is used in place of the simple Ising gate
$U_{\mathrm{zz}}(\theta)$ then the first and second-order effects of
the systematic error are eliminated.  In this manner, any number of
sequences designed for operations in $\SU{2}$ may be mapped into
sequences that compensate Ising gates, including the higher-order
Trotter-Suzuki sequences, which produce gates at an arbitrary level of
accuracy.  However, we note that since two-qubit gates occur so slowly,
the practical utility of very long sequences is not clear, especially
in systems where the two-qubit gate time is comparable to the qubit
coherence lifetime.  Substantial improvements in the minimum time
requirements may be possible with shaped pulse sequences or using
time-optimal control methods \cite{Lapert2011a}.

In this method, accurate one-qubit gates are used to transform
inaccurate Ising gates into a representation of $\SU{2}$.  Naturally,
it is unimportant which qubit among the pair is rotated to perform
this transformation, i.e., the subalgebra $\algb{j}' = \span\{ -\i
\H{1x}, -\i \H{zy}, -\i \H{zz} \}$ would serve just as well.  Given a
control with a systematic error and a perfect rotation operator that
transforms the control Hamiltonian $H_\mu$ to an independent
Hamiltonian $H_\nu$, it is possible to perform compensation if
$H_\mu$, and $H_\nu$ generate a representation of $\su{2}$
\cite{Tomita2010}.

\emph{Correcting simultaneous errors:}
Accurate single-qubit gates are required to compensate errors in the Ising coupling by
transforming Ising gates into the larger dynamical Lie
group.  If accurate single-qubit operations are not available, then
propagators of the form \eqn{eq:ising-rotation} can no longer be
reliably prepared.  However, if a compensating pulse sequence may be
implemented in place of each imperfect single-qubit rotation, then the
effect of this error may be reduced.
This procedure was used in section \ref{simultaneous1} to produce a
concatenated CORPSE and B2 sequence robust to simultaneous amplitude
and detuning errors.  A similar strategy can be employed to produce
accurate Ising gates in the presence of simultaneous spin-coupling
(two-qubit) and amplitude (one-qubit) systematic errors \cite{Tomita2010}.  Consider the
control system \eqn{eq:ising-hamiltonian} under the influence of these
two independent errors.  The imperfect propagator takes the form
\begin{eqnarray}
V(\u(t)) = U(\u(t) + \ep{A} \du_1(t) + \ep{J} \du_2(t)),
\end{eqnarray}
where $\ep{A} \du_1(t) = \ep{A} \ux(t) + \ep{A} \uy(t)$ is the
amplitude error of the single qubit controls, and $\ep{J} \du_2(t) =
\ep{J} \uu{zz}(t)$ is the error in the Ising coupling.  The systematic
error on the one-qubit controls $\ep{A} \du_1(t)$ complicates the
synthesis of compensated Ising gates, as unitary propagators of the
form \eqn{eq:ising-rotation} can no longer be reliably prepared, i.e.,
$\mathcal{M}(\theta,\phi) = R_1^\dagger(\phi(1+\ep{A}),0)
U_\mathrm{zz}(\theta(1+\ep{J})) R_1(\phi(1+\ep{A}),0) =
\mathcal{R}(\theta(1+\ep{J}),\phi) + \O(\ep{A}^2)$.  This difficulty
may be avoided if we use a B2 sequence to correct $\ep{A}$, before
correcting $\ep{J}$ using B2-J.  Let $U_\mathrm{B2}(\theta,\phi) = U_T
\otimes \Id + \O(\ep{A}^3)$ represent the propagator produced by a B2
sequence (see section \ref{simultaneous1}) for the target rotation
$U_T = R(\theta,\phi)$ on the first qubit.  The sequence B2-WJ is
\begin{eqnarray}
\mathcal{M}_{\text{B2-WJ}}(\theta,0) &=&  M_{\mathrm{B2}}^\dagger(\phi, 0)U_\mathrm{zz}(\pi(1+\ep{J})) M_{\mathrm{B2}}(\phi,0)  \: M_{\mathrm{B2}}^\dagger(3\phi, 0)U_\mathrm{zz}(2\pi(1+\ep{J})) M_{\mathrm{B2}}(3\phi,0) \: \times \nonumber \\ &\:& M_{\mathrm{B2}}^\dagger(\phi, 0)U_\mathrm{zz}(\pi(1+\ep{J})) M_{\mathrm{B2}}(\phi,0)\:U_\mathrm{zz}(\theta(1+\ep{J}))
\end{eqnarray}
This sequence replaces imperfect $R_1(\phi(1+\epsilon_A),0)$ pulses
with the compensated rotation produced by the B2 sequence.  When
$\epsilon_A = 0$, B2-WJ scales as $\O(\epsilon_J^3)$, and when
$\epsilon_J = 0$ the sequence scales as $O(\epsilon_A^3)$.

\begin{figure}
\begin{center}
\includegraphics{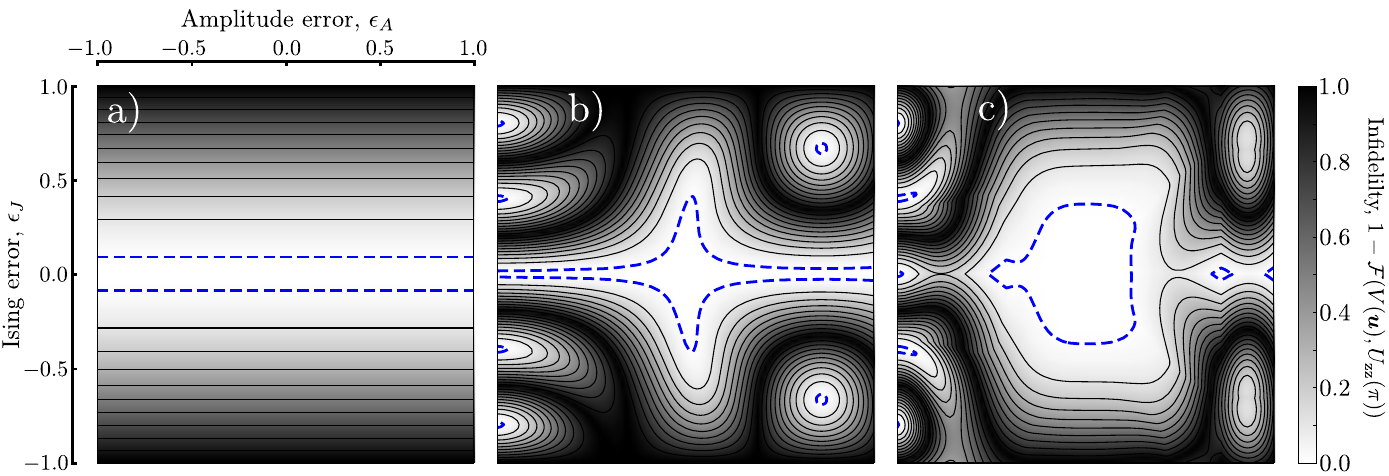}
\end{center}
\caption{Infidelity of a) B2-J b) B2-W and c) B2-WJ.  The target
    rotation is $U_T = U_{ZZ}(\pi,0)$. 
    The dashed contour corresponds to an infidelity of 0.01, while the
    remaining contours are plotted at 10\% intervals.}
\end{figure}


\subsubsection{Extension to $\SU{2^n}$:} This demonstrated for
two-qubits can be naturally extended to compensate operations on a
network of $n$ qubits with single qubit operations and Ising
couplings.  In this case the dynamical Lie algebra for this system is
$\SU{2^n}$.  Khaneja has identified a recursive Cartan decomposition
for this group that allows any $U \in \SU{2^n}$ to be written as a
product of single-qubit (local) and two-qubit (nonlocal) operations
\cite{Khaneja2001}.  Given a pair of sequences for both single-qubit
and two-qubit operations which compensate a particular error, it is
always possible to produce a sequence for any $U \in \SU{2^n}$ using
this decomposition.  In particular, imperfect Ising gates may be
replaced with a B2-J sequence, while imperfect single-qubit gates may
be corrected using the methods described in section \ref{SU2}. The
surprising result is that one needs only a single accurate control to
compensate an unlimited number of uncorrelated but systematic errors
\cite{Tomita2010}. In practice this concatenating scheme is expensive
for the whole system, but it points to a method for minimizing the
amount of calibration required since a few good controls can
compensate nearby errors.

\emph{Computation on subspaces:} Several interesting proposals involve
the encoding of logical qubits on subspaces of a larger Hilbert space,
which may offer certain advantages over other encoding schemes.  In
decoherence free subspace schemes, qubits are encoded on a subspace
which is decoupled from environmental noise sources \cite{Lidar2003,
  Weinstein2005, Storcz2005}.  Also, several theoretical proposals
involve the use of only two-qubit interactions to perform quantum
computation \cite{Childs2011}.  In these schemes, the control
algebra is chosen to be sufficiently large to allow universal
computation on a subspace.

An interesting question is whether gates applied to an encoded qubit
may be corrected by using a compensating sequence on the codespace
\cite{Tomita2010}.  Recall that any gate in $\SU{2^n}$ may be
decomposed as a product of single-qubit (encoded) and two-qubit gates;
it is sufficient to consider these cases individually.  We may reuse
the sequences described in section \ref{SU2} if the controls for the
encoded gates are distorted by a similar error model.  For example,
given a set of two controls with correlated systematic errors
$v_\mu(t) = (1 + \epsilon) u_\mu(t)$ and $v_\nu(t) = (1 + \epsilon)
u_\nu(t)$, it is possible to perform compensation if the control
Hamiltonians $H_\mu$ and $H_\nu$ generate a representation of
$\su{2}$. In Ref. \cite{Tomita2010}, it is shown that universal
subspace computation can be compensated if the two qubit Hamiltonians
are of the XY model, $H = \H{xx}+ \H{yy}$, but only single qubit
operations can be compensated if the two qubit couplings are of the
exchange type, $H = \H{xx} + \H{yy} + \H{zz}$.


\section{Conclusion and Perspectives}

Recent advances in quantum information and quantum control have
revitalized interest in compensating composite pulse sequences.
Specifically for the case of systematic control errors these
techniques offer a particularly resource efficient method for quantum
error reduction.  As quantum information processing experiments
continue to grow in both size and complexity, these methods are
expected to play an increasingly important role.

In this review, we have presented a unified picture of compensating
sequences based on control theoretic methods and a dynamic interaction
picture.  Our framework allows us to view each order of the error as a
path in the dynamical Lie algebra, highlighting the geometric
features. Correction of the first two orders has a natural geometric
interpretation: the path of errors must be closed and the signed area
enclosed by the path must be zero.


The geometric method helps illuminate the construction of arbitrarily
accurate composite pulses.  Currently arbitrarily accurate pulse
sequences are of limited use because as the length of the sequence
increases other noise sources become important.  For most experiments,
decoherence and random errors limit the fidelity of second-order
compensation sequences.  Our review of the Solovay-Kitaev pulses
resulted in a modest improvement of the sequence time by
implementing a new geometric construction.  Shorter sequences may be
achieved through numeric operation and continuous controls.

Finally, we note that CORPSE and related pulse sequences are similar
to dynamically corrected gates \cite{Khodjasteh2009}, in that both
remove coupling to an external field while performing an operation.
Combining the methods here with the developments in dynamically
corrected gates and dynamic decoupling could lead to pulse sequences
robust again environmental and control errors. These operations will
lower the initial error and in the end limit the resources required
for quantum error correction. This will ultimately determine the
feasibility of performing large quantum chemistry calculations on a
quantum computer \cite{Clark2009,Kassal2011}.

\section*{}
{\large{\bf Acknowledgments}}\\
\addcontentsline{toc}{section}{Acknowledgments} This work was
supported by the NSF-CCI on Quantum Information for Quantum Chemistry
(CHE-1037992) and by IARPA through ARO contract W911NF-10-1-0231.  JTM
acknowledges the support of a Georgia Tech Presidential Fellowship.

\section*{}
\addcontentsline{toc}{section}{References}

\end{document}